\documentclass{aa}

\usepackage{graphicx}
\usepackage{txfonts}
\usepackage{bm}

\usepackage{natbib}
\bibpunct{(}{)}{;}{a}{}{,}

\newcommand{\gcc}{\mbox{g cm$^{-3}$}}
\newcommand{\dync}{\mbox{dyn cm$^{-2}$}}
\newcommand{\req}[1]{Eq.~(\ref{#1})}
\newcommand{\dd}{\mathrm{d}}
\newcommand{\nb}{n_\mathrm{b}}
\newcommand{\nc}{n_\mathrm{cc}}
\newcommand{\nd}{n_\mathrm{d}}
\newcommand{\rhoc}{\rho_\mathrm{c}}
\newcommand{\Rc}{R_\mathrm{ch}}
\newcommand{\RWS}{R_\mathrm{WS}}

\begin{document}

\title{Analytical representations of 
unified equations of state \\
 for neutron-star matter\thanks{Our analytical fitting
expressions have been implemented in Fortran subroutines
that are publicly available
at the CDS via
anonymous ftp to cdsarc.u-strasbg.fr (130.79.128.5)
or via http://cdsweb.u-strasbg.fr/cgi-bin/qcat?J/A+A/,
or at
http://www.ioffe.ru/astro/NSG/BSk/}}
\titlerunning{Analytical equations of state for neutron-star
matter}
                                                         
\author{
A. Y. Potekhin\inst{1,2,3}\thanks{\email{palex@astro.ioffe.ru}}
\and
A. F. Fantina\inst{4}
\and
N. Chamel\inst{4}
\and
J. M. Pearson\inst{5}
\and
S. Goriely\inst{4}
}
\institute{
Centre de Recherche Astrophysique de Lyon, Universit\'e de
Lyon, Universit\'e Lyon 1, Observatoire de Lyon, Ecole Normale
Sup\'erieure de Lyon, CNRS, UMR 5574, 9 avenue Charles Andr\'e,
Saint Genis Laval F-69230, France
\and
Ioffe Physical-Technical Institute,
Politekhnicheskaya 26, St.~Petersburg 194021, Russia
\and
Isaac Newton Institute of Chile, 
         St.~Petersburg Branch, Russia
\and
Institut d'Astronomie et d'Astrophysique,
Universit\'e Libre de Bruxelles,
CP-226,
1050 Brussels, Belgium
\and
D\'epartement de Physique, 
Universit\'e de Montr\'eal, 
Montr\'eal (Qu\'ebec), H3C 3J7 Canada
}

\date{Received 15 April 2013 / Accepted 23 September 2013 /
Published 3 December 2013: \textbf{A\&A 560, A48 (2013)}}

%%%%%%%%%%%%%%%%%%%%%%%%%%%%%%%%%%%%%%%%%%%%%%%%%%%%%%%%%
\abstract
{An equation of state (EoS) of dense nuclear matter is
a prerequisite for studies of the structure and evolution of
compact stars. 
A unified EoS should describe the crust and the core of a
neutron star using the same physical model.
The Brussels-Montreal group has recently derived
a family of such EoSs based on the nuclear energy-density 
functional theory with generalized Skyrme effective forces
that have been fitted with great precision to essentially all the available
mass data. At the same time, these forces were constrained to reproduce 
microscopic calculations of homogeneous neutron matter based on realistic
two- and three-nucleon forces.
}{
We represent basic physical
characteristics of the latest Brussels-Montreal EoS models by
analytical expressions
to facilitate their inclusion in
astrophysical simulations.
}{
We consider three EoS models, which significantly differ by stiffness: BSk19, 
BSk20, and BSk21. For each of them we constructed two versions of the EoS
parametrization. In the first version, pressure $P$ and gravitational mass
density $\rho$ are given as functions of the baryon number density $\nb$. In
the second version, $P$, $\rho$, and $\nb$ are given as functions of 
pseudo-enthalpy, which is useful for two-dimensional calculations of stationary
rotating configurations of neutron stars. In addition to the
EoS, we derived
analytical expressions for several related quantities that are required in 
neutron-star simulations: number fractions of electrons and muons in the 
stellar core, nucleon numbers per nucleus in the inner crust,
and equivalent radii 
and shape parameters of the nuclei in the inner crust.
}{
We obtain analytical representations for the basic
characteristics of the models of cold dense matter,
which are most important for studies of neutron
stars. We demonstrate the usability of our results by
applying them to calculations of neutron-star mass-radius
relations, maximum and minimum masses, thresholds of direct
Urca processes, and the electron conductivity in
the neutron-star crust.
}
{}


\maketitle

%%%%%%%%%%%%%%%%%%%%%%%%%%%%%%%%%%
\section{Introduction}
\label{sect:introd}
%%%%%%%%%%%%%%%%%%%%%%%%%%%%%%%%%

The equation of state (EoS) of dense matter is crucial as
input for neutron-star structure calculations. Usually,
neutron-star matter is strongly degenerate,  and therefore
the EoS is barotropic (i.e., the pressure is
temperature-independent), except for the outermost envelopes
(a few meters thick).

A \emph{unified EoS} is based on a single effective nuclear
Hamiltonian and is valid in  all regions of the neutron-star
interior. For unified EoSs, the transitions between the outer
crust and the inner crust, and between the inner and
the core are treated consistently using the same physical
model \citep[e.g.,][]{DouchinHaensel00,Pearson-ea12}.  Other
(nonunified) EoSs consist of crust and core segments
obtained using different models. The crust-core interface
there has no physical meaning, and both segments are joined
using an ad hoc matching procedure. This generally leads to
thermodynamic inconsistencies, which can  manifest
themselves by the occurrence of spurious instabilities in 
neutron-star dynamical simulations. Even if the transitions
are treated  using a unified approach, instabilities might
still arise due to numerical  errors. In addition, realistic
EoSs are generally calculated only for specific  densities
and/or pressures. These limitations can be circumvented  by
using analytical representations of the EoSs.  
In this paper we construct analytical representations of three recent 
unified EoSs for cold catalyzed nuclear matter developed by the Brussels-Montreal 
group: BSk19, BSk20, and BSk21~\citep{Goriely-ea10,PearsonGC11,Pearson-ea12}.

We follow the approach developed by \citet{HP04}, who
constructed analytical representations for the previous
unified EoS models FPS \citep{Pandharipande89} and SLy4
\citep{DouchinHaensel01}. In addition, we present analytical
parametrizations of the composition of the crust
and number fractions of leptons and nucleons in the
core of a neutron star. We adopt the ``minimal model''
\citep[e.g.,][]{HPY}, which means we assume the nucleon-lepton
matter without exotic particles.

The composition of the neutron-star crust can depend on its
formation history. For example, if the star experienced
accretion, then the crust could be formed by the nuclear
transformations that accompany gradual density increase
under the weight of the newly accreted matter \citep{HZ90}.
It differs from the matter in beta-equilibrium (cold
catalyzed matter) that constitutes the crust soon after the
birth of a neutron star. Moreover, the nonequilibrium
composition of an accreted crust can also depend on the
composition of the ashes of accretion-induced thermonuclear
burning \citep{HZ03}. Here, however, we do not consider
these complications, but focus on the cold catalyzed matter.

In Sec.~\ref{sect:propeos} we briefly review the nuclear models 
used to construct the Brussels-Montreal EoSs, including a discussion 
of constraints coming from nuclear physics experiments and many-body 
calculations.
In Sect.~\ref{sect:EoS} we present a set of fully analytical
approximations to the EoSs in the crust and the core of a
neutron star. This set includes pressure $P$ and baryon
number density $\nb$ as  functions of gravitational mass
density $\rho$,  the inverse function $\rho(\nb)$, and a fit
of $\rho$ as a function of the pseudo-enthalpy, which is a
particularly convenient independent variable for simulations
of rapidly rotating neutron stars \citep{Bonazzola-ea93}. In
Sect.~\ref{sect:frac} we give analytical approximations to
number fractions of electrons and muons in the core and the
inner crust of a neutron star as functions  of $\nb$.
Based on the one-dimensional
approximation of the nuclear shapes suggested by
\citet{Onsi-ea08}, we describe the shapes of the nuclei in
the inner crust as fully analytical functions of two
arguments, the radial coordinate in a Wigner-Seitz cell $r$
and the mean baryon density $\nb$. We also present effective
proton and neutron sizes of the nuclei in the inner crust as
analytical functions of $\nb$. In Sect.~\ref{sect:cond} we
consider an application of the results to the calculation of
electron conductivity in the stellar crust. The impact of
the analytical representation  of the EoSs on neutron-star
structure is studied in  Sect.~\ref{sec:fit_ns}.  Concluding
remarks are given in Sect.~\ref{sect:concl}.

%%%%%%%%%%%%%%%%%%%%%%%%%%%%%%%%%%%%%%%%%%%%%%%%%%%%%%%%%%%%%%%%%%%%%%%%

\section{The unified Brussels-Montreal EoSs}
\label{sect:propeos}

%%%%%%%%%%%%%%%%%%%%%%%%%%%%%%%%%%%%%%%%%%
The unified Brussels-Montreal EoSs that we consider here are based on the 
nuclear energy-density functionals (EDFs), labeled BSk19,
BSk20, and 
BSk21, respectively. These EDFs were derived from generalized Skyrme 
interactions, supplemented with microscopic contact pairing interactions,
a phenomenological Wigner term and correction terms for the collective 
energy~\citep{Goriely-ea10,Chamel10}. Calculating the nuclear energy with
the Hartree-Fock-Bogoliubov (HFB) method, the EDFs were fitted to the 2149 
measured masses of atomic nuclei with proton number $Z\geq 8$ and neutron 
number $N\geq 8$ from the 2003 Atomic Mass Evaluation~\citep{audi2003} with an 
root mean square (rms) deviation as low as 0.58 MeV. In making these fits the Skyrme part of the
EDFs were simultaneously constrained to fit the zero-temperature EoS of 
homogeneous neutron matter (NeuM), as determined by many-body calculations with 
realistic two- and three-nucleon forces. Actually, several realistic 
calculations of the EoS of NeuM have been made, and while they all agree fairly
closely at nuclear and subnuclear densities, at the much higher densities that 
can be encountered towards the center of neutron stars they differ greatly in 
their stiffness, and there are very few data, either observational or 
experimental, to distinguish between the different possibilities. 
It is in 
this way that the three different functionals were
constructed, as follows. 
The BSk19 EDF was constrained to the soft \citet{FP81} EoS obtained from the 
realistic Urbana v14 nucleon-nucleon force with the three-body force TNI, the 
BSk20 EDF was fitted to the \citet{APR} EoS labeled ``A18 +
$\delta v$ + UIX'',
and the BSk21 EDF was adjusted to the stiff EoS labeled
``V18'' in 
\citet{LiSchulze}.

These NeuM constraints make the EDFs BSk19--21 suitable for application
to the neutron-rich environments encountered in many different astrophysical
situations, and our making three different such EDFs available reflects the
current lack of knowledge of the high-density behavior of dense matter.
In addition, these three EDFs were also constrained to reproduce several other
properties of homogeneous nuclear matter as obtained from 
many-body calculations using realistic two- and three-nucleon interactions; 
among those, the ratio of the isoscalar effective mass
$m^*_\mathrm{s}$ to bare
nucleon mass $m$ in symmetric 
nuclear matter at saturation was set to the realistic value of 0.8, and 
all three EDFs predict a neutron effective mass that is
higher than the 
proton effective mass in neutron-rich matter, as found both experimentally and
from  microscopic calculations. Various properties of 
the BSk19, BSk20, and BSk21 EDFs are summarized in Table~\ref{tab:bskprop} 
\citep{Goriely-ea10}, namely: the rms deviations
to the 2149 measured atomic masses
$\sigma(m_\mathrm{atom})$ and to the 
782 measured charge radii $\sigma(\Rc)$, the energy per nucleon of symmetric nuclear matter at saturation density 
$a_v$, the baryon number density at saturation $n_{{\rm b},0}$, the incompressibility of symmetric 
nuclear matter at saturation $K_v$ (which was required to fall in the experimental range $240\pm10$~MeV, 
according to \citealt{colo04}), the symmetry energy coefficient $J$ and its slope $L$, the isospin 
compressibility $K_\tau$, the isoscalar and isovector
effective masses $m^*_\mathrm{s}$ and $m^*_\mathrm{v}$ relative to the bare
nucleon mass $m$, and the limiting baryonic 
density $n_\mathrm{caus}$
after which the EoSs of neutron-star matter violate causality. The last line indicates the 
NeuM EoS to which each EDF was fitted.

\begin{table}
\centering
\caption[]{Properties of the Skyrme forces
BSk19, BSk20, and BSk21 \citep{Goriely-ea10}. 
See text for details.}
\label{tab:bskprop}
\begin{tabular}{r|ccc}
\hline\hline\rule[-1.4ex]{0pt}{4.3ex}
  & BSk19 & BSk20 & BSk21  \\
\hline\rule{0pt}{2.7ex}
$\sigma(m_\mathrm{atom})$~[MeV] & 0.583 & 0.583 & 0.577 \\
$\sigma(\Rc)$~[fm] & 0.0283 & 0.0274 & 0.0270 \\
$a_v$~[MeV]  & $-16.078$ & $-16.080$ & $-16.053$ \\
$n_{{\rm b},0}$~[fm$^{-3}$] & 0.1596 & 0.1596 & 0.1582 \\
$K_v$~[MeV] & 237.3 & 241.4 & 245.8 \\
$J$~[MeV] & 30.0 & 30.0 & 30.0 \\
$L$~[MeV] & 31.9 & 37.4 & 46.6 \\
$K_\tau$~[MeV] & $-342.8$ & $-317.1$ & $-264.6$ \\
$m^*_\mathrm{s}/m$ & 0.80 & 0.80 & 0.80 \\
$m^*_\mathrm{v}/m$ & 0.61 & 0.65 & 0.71 \\
$n_{\rm caus}$~[fm$^{-3}$] & 1.45 & 0.98 & 0.99 \\
\hline
NeuM & FP$^\mathrm{a}$ & APR$^\mathrm{b}$ & LS$^\mathrm{c}$ \\
\hline\hline
\end{tabular}
\tablebib{(a)~\citet{FP81};
           (b)~\citet{APR};
           (c)~\citet{LiSchulze}. }
\end{table}

The Brussels-Montreal EDFs BSk19, BSk20, and BSk21 were used to compute the EoS 
of all regions of a neutron star. Following the BPS model \citep{bps1971}, the 
outer crust was assumed to consist of fully ionized atoms arranged in a 
body-centered cubic lattice at zero temperature. The EoSs of the outer crust 
were calculated using either experimental atomic masses when available or 
theoretical masses obtained from the HFB mass models constructed with the 
BSk19, BSk20, and BSk21 EDFs, as appropriate 
\citep[see][for details]{PearsonGC11}. For the inner crust, where 
neutron-proton clusters coexist with free neutrons, the kinetic-energy part of
the appropriate EDF was calculated using the semi-classical extended 
Thomas-Fermi method with 
proton quantum shell corrections added via the Strutinsky integral theorem; 
neutron shell effects, which are known to be much smaller, were neglected 
\citep[see][for details]{Pearson-ea12}. This method is a computationally very 
fast approximation to the full Hartree-Fock equations. In order to further 
reduce the computations, nuclear clusters were assumed to be spherical, and 
parametrized nucleon distributions were introduced. Finally, the electrostatic 
energy was calculated using the Wigner-Seitz approximation, and pairing effects
were neglected. The overall resulting errors in the EoS of the inner crust were
found to be about 5\% at the neutron-drip point. The EoSs of the core, assumed
to consist of homogeneous beta-equilibrated matter made of nucleons and leptons 
(electrons and muons), were calculated essentially analytically from the 
adopted EDF \citep{Goriely-ea10}. 

In a recent paper, \citet{dutra2012} analyzed 
240 Skyrme parametrizations, including
BSk EDFs, by comparing their predicted 
properties of symmetric nuclear matter and pure NeuM to some 
empirical constraints. On this basis, these authors rejected
BSk19--21 EDFs
(along with all but five of the 240 parametrizations that they considered). We
now argue that this rejection is unjustified. In the first 
place, BSk19--21 EDFs are claimed to be incompatible with the constraint
labeled ``SM3" by \citet{dutra2012}, a constraint on the EoS of symmetric nuclear
matter obtained from the analysis of particle-flow measurements in Au-Au
collisions \citep{danielewicz2002} that can be represented by a band in the
plot of pressure vs.{} density. Now the pressures calculated with the EDFs
BSk19--21 fall within this band over 80\% of the density range and never 
deviate by more than about 20\% from those inferred by 
\citet{danielewicz2002} (see, e.g., Fig. 3 of \citealt{Chamel-ea11}). Nevertheless,
\citet{dutra2012} reject these functionals on the grounds that the calculated EoSs
do not fall within the band of \citet{danielewicz2002} over 95\% of the density range. 
This criterion is quite arbitrary, and without any
sound statistical foundation. 
In addition, it has to be noted that the interpretation 
by \citet{danielewicz2002} of the raw experimental data is subject to
uncertainties of two different kinds of model dependence: i) the transport
model that determines the particle flow in a given collision experiment;
ii) extrapolation from the charge-asymmetric Au + Au system ($(N - Z)/A \approx
20\% $) to symmetric nuclear matter. 
As demonstrated by \citet{Gale-ea90}, the flow is generated
during the early nonequilibrium stages of the collision,
whereas the interpretation in terms of the EoS by 
\citet{danielewicz2002} is an equilibrium version of a
simplified momentum-dependent interaction.
Viewed in this light, the deviations of 
the pressures obtained with BSk19--21 from the values of
\citet{danielewicz2002} are altogether insignificant.

The EDFs BSk19-21 are also found to violate the NeuM constraint 
``PNM2" by \citet{dutra2012}. However, this
constraint was actually obtained by \citet{danielewicz2002} from the symmetric 
nuclear matter constraint ``SM3" using a simple parametrization for the 
symmetry energy. For this reason,
the constraint ``PNM2" is even less meaningful than the constraint ``SM3".

\begin{figure}[h]
\begin{center}
\includegraphics[width=8cm]{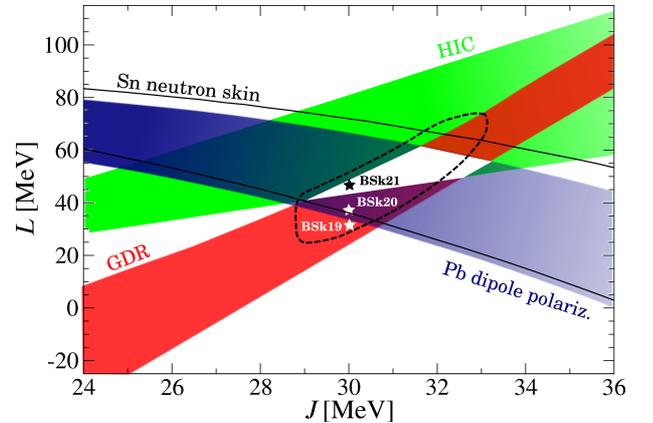}
\end{center}
\caption{Experimental constraints on the symmetry energy parameters (see text for details), 
taken from \citet{lattimerlim2013}. The dashed line represents the constraint obtained from fitting 
experimental nuclear masses using the Brussels-Montreal HFB models with a  root-mean-square deviation 
below 0.84 MeV (best fits are for $J=30$~MeV). Star
symbols correspond to the models BSk19, BSk20, and BSk21.}
\label{fig_jl}
\end{figure}

\citet{dutra2012} also studied the density dependence of the symmetry energy. 
In particular, they found that the coefficient $L$ obtained
with BSk19 and BSk20 EDFs falls 
outside the range of empirical values determined by
\citet{chen2010} thus violating the constraints labeled
``MIX2'' and ``MIX5''. However, the situation 
regarding the symmetry energy still remains a matter of debate, different 
experiments and/or theoretical calculations leading to different and 
sometimes contradicting predictions, as shown e.g. in Fig.~12 of 
\citet{lattimer2012} \citep[see also][Sect.~IIIB]{tsang2012, Goriely-ea10}. 
Different Skyrme EDFs can thus be selected depending on the experiments or 
many-body calculations. For example, the values of $L$
obtained with BSk19--21 EDFs 
are all compatible with the range of values $L=55\pm25$~MeV deduced by 
\citet{centelles2009} and \citet{warda2009} from measurements of the 
neutron-skin thickness in nuclei. The $L$ coefficients obtained with BSk20 and 
BSk21 are also in agreement with the values $36$~MeV$< L < 55$~MeV found by 
\citet{steinergandolfi2012} combining quantum Monte Carlo calculations with 
neutron-star mass and radius measurements. Incidentally, this latter constraint
would rule out three of the only five functionals that survive the analysis
of \citet{dutra2012}. Various other constraints on $J$ and
$L$ obtained from the analysis 
of different experiments \citep{tsang2012,lattimerlim2013}
have been summarized in Fig.~\ref{fig_jl}.
The figure shows the constraint deduced by \citet{tsang2009} from heavy-ion collisions 
(HIC), that derived by \citet{chen2010} from measurement of the neutron skin thickness in 
tin isotopes,  and that obtained from the analysis of giant dipole resonance (GDR) 
\citep{trippa2008, lattimerlim2013} and the electric dipole polarizability~\citep{piekarewicz2012} 
in $^{208}$Pb. For comparison, the constraint obtained from
our HFB nuclear mass models BSk9--26
 \citep[][and references therein]{Goriely-ea13,Pearson2009}
as well as unpublished ones are also shown. Values of $J$ and $L$ have also been extracted from 
pygmy dipole resonances \citep{carbone2010} and isobaric analogue states \citep{danielewiczlee2009}. 
However, we have not included them in the figure because of large experimental and theoretical 
uncertainties \citep[see, e.g.,][]{reinhardnazarewicz2010, daoutidisgoriely2011}. The 
HFB mass model not only agrees with GDR, neutron skin and dipole polarizability data, but also 
provides relatively strict constraints on the $J$ an $L$ values.

\citet{dutra2012} also  point out that the values of $K_\tau$ obtained for the EDFs 
BSk19, BSk20, and BSk21 are incompatible with the range
$-760$~ MeV $< K_\tau < -372$~ MeV that they extract from experimental data 
using a liquid-drop like approach (the constraint labeled
``MIX3''). On the other hand, the values of $K_\tau$ 
obtained with BSk19, BSk20, and BSk21 are all compatible with the range of 
values $-370\pm120$~MeV inferred from isospin diffusion in heavy-ion collisions
by \citet{chen2009}. Just as for the $L$ coefficient, the uncertainties in 
$K_\tau$ still remain very large \citep[see e.g. Sect.~IIIC in][for a 
thorough discussion]{Goriely-ea10}. 

Thus the rejection of BSk19--21 EDFs by \citet{dutra2012} is
ungrounded. On the contrary, these EDFs are well adapted
to a unified treatment of all parts of neutron stars: the outer crust, the 
inner crust and the core. 
The relevance of these EDFs to the core of neutron stars arises not 
only from their fit to the EoS of NeuM but also from the good 
agreement between their predicted EoS of symmetric nuclear matter and 
realistic calculations, which implies that they take correct account of 
the presence of protons. Since these EDFs were fitted to nuclear masses, 
they also take account of inhomogeneities, and thus are appropriate for 
the calculation of the inner crust of neutron stars. As for the outer 
crust, its properties are determined entirely by the mass tables that 
have been generated for the appropriate EDFs by \citet{PearsonGC11}.

%%%%%%%%%%%%%%%%%%%%%%%%%%%%%%%%%%%%%%%%%%%%%%%%%%%%%%%%%%%%%%%%%%%%%%%%
\section{Analytical representations of the EoS}
\label{sect:EoS}

%%%%%%%%%%%%%%%%%%%%%%%%%%%%%%%%%%%%%%%%%%
\subsection{Preliminary remarks}
\label{sect:prelim}

The first law of thermodynamics for a barotropic EoS implies the
relation (see, e.g., \citealt{HaenselProsz82})
$
P(\nb)=\nb^2 c^2 \dd(\rho/\nb)/\dd\nb,
$
which can be also used in the integral
forms:
%%%%%%%
\begin{equation}
\frac{\rho(\nb)}{\nb} = \frac{\rho_\mathrm{s}}{n_\mathrm{s}} +
\int_{n_\mathrm{s}}^{\nb} {P(\nb')\over {\nb'}^2 c^2}\,
\dd \nb',
\quad
\ln\left(\frac{\nb}{n_\mathrm{s}}\right) = 
 \int_{\rho_\mathrm{s}}^\rho
 \frac{c^2\,\dd\rho'}{P(\rho') + \rho' c^2} ,
\label{Integral}
\end{equation}
where $\rho_\mathrm{s}$ and $n_\mathrm{s}$ are the values of
$\rho$ and $\nb$ at the neutron-star surface. In the present
paper we put $\rho_{\rm s}$ equal to the density of $^{56}{\rm
Fe}$ at zero pressure and zero temperature, $\rho_{\rm
s}=7.86$ \gcc. One of the advantages of an analytical
representation of the EoS is that it allows one to fulfill
the relations (\ref{Integral}) precisely.

There are three qualitatively different domains of the interior of a neutron 
star: the outer crust, which consists of electrons and
atomic nuclei; the inner crust, which consists of electrons, neutron-proton 
clusters, and ``free'' neutrons; and the core, which
contains electrons, neutrons, protons, muons, and possibly other particles 
(see, e.g., \citealt{HPY} for review and references). In addition,
there can be density discontinuities at the interfaces
between layers containing different nuclei in the crust. An
approximation of the EoS by a fully analytical function
neglects these small discontinuities. However, the different
character of the EoS in the three major domains is
reflected by the complexity of our fit, which consists of
several fractional-polynomial parts, matched together with
the use of the function
$
    (\mathrm{e}^x+1)^{-1}.
$

For $\rho<10^6$ \gcc{} the BSk EoSs are inadequate, primarily because atoms are 
not completely ionized, and thermal effects become non-negligible. 
The temperature dependence can be
roughly described as
\citep{HP04} $P=P_\mathrm{fit}+P_0$, where $P_\mathrm{fit}$
is given by the fitting function presented below, and
$P_0=A(T)\,\rho$ provides a low-density continuation. For
example, if the outer envelope consists of fully ionized
iron,  then
$A(T)\approx4\times10^7\,T$~K$^{-1}$~(cm/s)$^2$.
Partial ionization decreases $A(T)$: for example, at
$T=10^7$~K the best interpolation to the OPAL EoS
\citep{OPAL-EoS} is given by
$A=3.5\times10^{14}$~(cm/s)$^2$ (that is, 14\% smaller
than for the fully ionized ideal gas).

We constructed analytical parametrizations for pressure
$P$, gravitational mass density $\rho$, and baryon number
density $\nb$ as functions of $\rho$,
$\nb$, or pseudo-enthalpy  
\begin{equation}
H = \int_0^P \frac{\dd P'
         }{ \rho (P')\,c^2 +P'}.
\label{H.def}
\end{equation}
The latter quantity is a convenient variable for models of
rotating stars in General Relativity
(see \citealt{Sterg03}, for review).

%%%%%%%%%%%%%%%%%%%%%%%%%%%%%%%%%%%%%%%%%%
\subsection{Pressure as a function of density}

We introduce the variables $\xi=\log(\rho/\textrm{g cm}^{-3})$ and
$\zeta = \log(P/\dync)$. Here and hereafter, $\log$ denotes 
$\log_{10}$, while the natural logarithm is denoted by
$\ln$.
Our parametrization of $P(\rho)$ reads
\begin{eqnarray}
  \zeta &=&
    \frac{a_1+a_2\xi+a_3\xi^3}{1+a_4\,\xi}\,
        \left\{\exp\left[a_5(\xi-a_6)\right]+1\right\}^{-1}
\nonumber\\&&
     + (a_7+a_8\xi)\,
        \left\{\exp\left[a_9(a_6-\xi)\right]+1\right\}^{-1}
\nonumber\\&&
     + (a_{10}+a_{11}\xi)\,
\left\{\exp\left[a_{12}(a_{13}-\xi)\right]+1\right\}^{-1}
\nonumber\\&&
     + (a_{14}+a_{15}\xi)\,
\left\{\exp\left[a_{16}(a_{17}-\xi)\right]+1\right\}^{-1}
\nonumber\\&&
     + \frac{a_{18}}{1+ [a_{19}\,(\xi-a_{20})]^2}
     + \frac{a_{21}}{1+ [a_{22}\,(\xi-a_{23})]^2}
     .
\label{fit.P}
\end{eqnarray}
The parameters $a_i$ are given in
Table~\ref{tab:fit.P}. The typical fit error of $P$ is
$\approx1$\% for $\xi\gtrsim6$. The maximum error is 3.7\%
at $\xi=9.51$; it is determined
by the jumps at the interfaces between layers containing
different nuclides in the tabulated EoS. The fit (\ref{fit.P})
smoothly interpolates across these jumps.

\begin{table}
\centering
\caption[]{Parameters of \req{fit.P}.}
\label{tab:fit.P}
\begin{tabular}{r|ccc}
\hline\hline\rule[-1.4ex]{0pt}{4.3ex}
$i$ & \multicolumn{3}{c}{$a_i$}\\
  & BSk19 & BSk20 & BSk21  \\
\hline\rule{0pt}{2.7ex}
1  & 3.916 &     4.078 &   4.857 \\
2  & 7.701 &     7.587 &   6.981 \\
3  & 0.00858 &   0.00839 & 0.00706 \\
4  & 0.22114 &   0.21695 & 0.19351 \\
5  & 3.269 &     3.614 &   4.085 \\
6  & 11.964 &    11.942 &  12.065 \\
7  & 13.349 &    13.751 &  10.521 \\
8  & 1.3683 &    1.3373 &  1.5905 \\
9  &  3.254 &    3.606 &   4.104 \\
10 &  $-12.953$ &  $-22.996$ & $-28.726$ \\
11 &  0.9237 &   1.6229 &  2.0845 \\
12 &  6.20 &     4.88 &    4.89 \\
13 &  14.383 &   14.274 &  14.302 \\
14 &  16.693 &   23.560 &  22.881 \\
15 &  $-1.0514$ &  $-1.5564$ & $-1.7690$ \\
16 &  2.486 &    2.095 &   0.989 \\
17 & 15.362 &    15.294 &  15.313 \\
18 & 0.085 &     0.084 &   0.091 \\
19 & 6.23 &      6.36 &    4.68 \\
20 & 11.68 &     11.67 &   11.65 \\
21 & $-0.029$ &    $-0.042$ &  $-0.086$ \\
22 & 20.1 &      14.8 &    10.0 \\
23 & 14.19 &     14.18 &   14.15 \\
\hline\hline
\end{tabular}
\end{table}

Compared to Eq.~(14) of \citet{HP04},
\req{fit.P} contains
additional terms in the last line with coefficients
$a_{18}$\,--\,$a_{23}$. These terms improve the fit near the
boundaries between the outer and inner crust and between the
crust and the core, where the slope of $P(\rho)$ sharply
changes. In SLy4 the analogous changes were less abrupt.
In the case of BSk models, however, the residuals of the fit
without these additional terms may reach about 10\%
\citep{Fantina-ea12}.

\begin{figure}
\begin{center}
\includegraphics[width=\columnwidth]{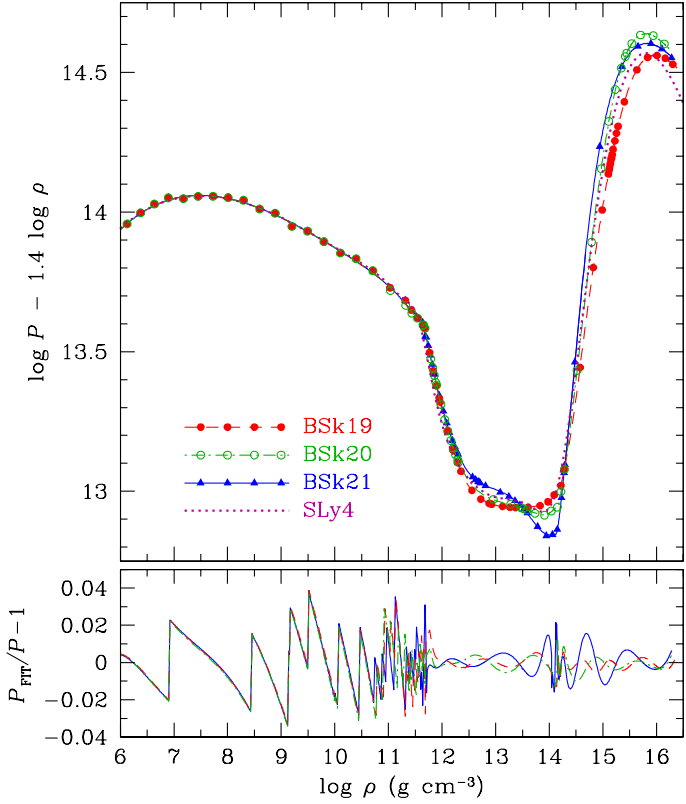}
\end{center}
\caption{Comparison of the data and fits for the pressure as
a function of mass density for the EoS models BSk19, BSk20,
and BSk21. Upper panel:
rarefied tabular data (symbols) and the fit (\ref{fit.P}) (lines);
lower panel: relative difference between the data and fit.
Filled dots and dashed
lines: BSk19; open circles and dot-dashed lines: BSk20;
filled triangles and solid lines: BSk21.
For comparison, the dotted line in the upper panel
reproduces the fit to the EoS SLy4 \citep{HP04}.
}
\label{fig:Pfit}
\end{figure}

In Fig.~\ref{fig:Pfit} we compare the EoSs BSk19, BSk20, and
BSk21 with their analytical representations. Symbols  on the
upper panel show the data, and lines show the fit. For
comparison, the additional dotted line represents SLy4 EoS
according to the fit of \citet{HP04}. In order to make the
differences between different EoSs visible, we plot  the
function $\zeta - 1.4 \xi$ (cf.~Fig.~4 of \citealp{HP04}).
The data points in the figure are rarefied, i.e., we show
only a small fraction of  all numerical data used for the
fitting. The lower panel of Fig.~\ref{fig:Pfit} shows the 
difference between the tabulated and fitted EoSs and thus
illustrates the accuracy of \req{fit.P}.

%%%%%%%%%%%%%%%%%%%%%%%%%%%%%%%%%%%%%%%%%%
\subsection{Mass density versus number density}
\label{sect:rho-n}

In some applications, it may be convenient to use $\nb$ as
independent variable, and treat $\rho$ and $P$ as functions
of $\nb$. For this purpose we construct a fit to the
deviation from the linear law $\rho={\nb m_\mathrm{u}}$,
where $m_\mathrm{u} = 1.66\times10^{-24}$~g is the 
atomic mass unit:
\begin{equation}
   \Delta_\rho \equiv \frac{\rho}{\nb m_\mathrm{u}} -1.
\label{Delta}
\end{equation}
In view of the relation (see, e.g., \citealt{HP04})
\begin{equation}
H=\ln(h/h_\mathrm{s}),
\label{H-h}
\end{equation}
where $h=(\rho c^2 +P)/\nb$ is the zero-$T$ enthalpy per baryon
and $h_\mathrm{s}$ is the value of $h$ at the stellar surface,
$\Delta_\rho$ must be consistent with \req{fit.P}.
 In order to fulfill \req{H-h} as closely as possible,
 we first calculate $H(\rho)$
  using Eqs.~(\ref{H.def}) and (\ref{fit.P}),
and then refine the values of $\nb$ using \req{H-h}.
Our fit to the result is
\begin{equation}
   \Delta_\rho  = 
     (1-f_1)\,\frac{b_1 n^{b_2}+b_3 \sqrt{n}}{(1+b_4 n)^2}
     +  
   f_1\,  \frac{n}{b_5+ b_6 n^{b_7}},
\label{fit.rho-n}
\end{equation}
where $ f_1 \equiv 
\left[\exp(1.1\log n + b_8)+1\right]^{-1}$,
$n=\nb/\textrm{fm}^{-3}$, 
and parameters $b_i$ are given in
Table~\ref{tab:fit.rho-n}. The typical error of \req{fit.rho-n} for
$\Delta_\rho$ is (1--2)\%, and the maximum relative error
of 4.2\% is at the low end of the fitted density range,
$\min(\rho)=10^6$ \gcc{}
(but at such low densities $\Delta_\rho$ is itself
negligible).

\begin{table}
\centering
\caption[]{Parameters of \req{fit.rho-n}.}
\label{tab:fit.rho-n}
\begin{tabular}{r|ccc}
\hline\hline\rule[-1.4ex]{0pt}{4.3ex}
$i$ & \multicolumn{3}{c}{$b_i$}\\
  & BSk19 & BSk20 & BSk21  \\
\hline\rule{0pt}{2.7ex}
1  &   0.259 &            0.632 &             3.85 \\
2  &    2.30 &           2.71 &              3.19 \\
3  & 0.0339 &            0.0352 &            0.0436 \\
4  & 0.1527 &            0.383 &             1.99 \\
5  & $1.085\times10^{-5}$ & $1.087\times10^{-5}$ & $1.075\times10^{-5}$ \\
6  & 3.50 &              3.51 &              3.44 \\
7  & 0.6165 &            0.6167 &            0.6154 \\
8  & 3.487 &             3.4815 &             3.531 \\
\hline\hline
\end{tabular}
\end{table}

\begin{table}
\centering
\caption[]{Parameters of \req{fit.n-rho}.}
\label{tab:fit.n-rho}
\begin{tabular}{r|ccc}
\hline\hline\rule[-1.4ex]{0pt}{4.3ex}
$i$ & \multicolumn{3}{c}{$c_i$}\\
  & BSk19 & BSk20 & BSk21  \\
\hline\rule{0pt}{2.7ex}
1  & 0.378 &     0.152 &     0.085 \\
2  & 1.28 &      1.02 &      0.802 \\
3  & 17.20 &     10.26 &     16.35 \\
4  & 3.844 &     3.691 &     3.634 \\
5  & 3.778 &     2.586 &     2.931 \\
6  & $1.071\times10^{-5}$ & $1.067\times10^{-5}$ & $1.050\times10^{-5}$ \\
7  & 3.287 &     3.255 &     3.187 \\
8  & 0.6130 &    0.6123 &    0.6110 \\
9  & 12.0741 &   12.0570 &   12.0190 \\
\hline\hline
\end{tabular}
\end{table}

The inverse fit $\nb(\rho)$ is given by the formula
\begin{equation}
   \frac{\tilde{\rho}}{n} = 1+
     (1-f_2)\,\frac{c_1 \tilde{\rho}^{c_2}
      + c_3 \tilde{\rho}^{c_4}}{(1+c_5 \tilde{\rho})^3}\,
     +  
    \frac{\tilde{\rho}}{c_6 + c_7\,\tilde{\rho}^{c_8}}\,f_2
 ,
\label{fit.n-rho}
\end{equation}
where $f_2 \equiv \left[\exp(\xi-c_9)+1\right]^{-1}$, the
dimensionless argument is defined as $\tilde{\rho} \equiv
(\rho/m_\mathrm{u})$~fm$^3 = \rho/(1.66\times10^{15}~\gcc)$,
and parameters $c_i$ are given in Table~\ref{tab:fit.n-rho}.
As well as  \req{fit.rho-n}, the fit (\ref{fit.n-rho}) also
minimizes $\Delta_\rho$, and its residuals are similar: the
average error is less than 2\%, and the maximum relative
error is 3.8\% at the lower end of the fitted density range.

%%%%%%%%%%%%%%%%%%%%%%%%%%%%%%%%%%%%%%%%%%
\subsection{Density as a function of pseudo-enthalpy}

As noted in Sect.~\ref{sect:prelim}, analytical expressions
of $\rho$ and $P$ as functions of the pseudo-enthalpy $H$
are expected to be useful for numerical simulations of
rotating stars. In order to achieve the  maximal consistency
of our parametrizations, we first calculate $H(\rho)$ using
Eqs.~(\ref{H.def}) and (\ref{fit.P}), and then parametrize
the result (cf.~Sect.~\ref{sect:rho-n}). The best fit reads
\begin{eqnarray}
  \xi &=& \left[ 2.367+ \frac{21.84\,\eta^{0.1843}}{1+0.7\eta}
       \right] f_3
\nonumber\\&&
     +  \frac{d_1+d_2\log\eta+(d_3+d_4\log\eta)(d_5\eta)^{d_{10}}
       }{
        1+d_6\eta+(d_5\eta)^{d_{10}}}\,(1-f_3)
\nonumber\\&&
     +  d_7\,\left\{\exp\left[d_8
                 (d_9-\log\eta)\right]+1\right\}^{-1},
\label{fit.rho-h}
\end{eqnarray}
where  $\eta\equiv \mathrm{e}^H-1$,
$f_3 \equiv
\left[\exp(84\log\eta+167.2)+1\right]^{-1}$,
and the parameters $d_i$ are given in Table~\ref{tab:fit.rho-h}.

\begin{table}
\centering
\caption[]{Parameters of \req{fit.rho-h}.}
\label{tab:fit.rho-h}
\begin{tabular}{r|ccc}
\hline\hline\rule[-1.4ex]{0pt}{4.3ex}
$i$ & \multicolumn{3}{c}{$d_i$}\\
  & BSk19 & BSk20 & BSk21  \\
\hline\rule{0pt}{2.7ex}
1  & 93.650 & 81.631 & 63.150 \\
2  & 36.893 & 31.583 & 23.484 \\
3  & 15.450 & 15.310 & 15.226 \\
4  & 0.672 &   0.594 &   0.571 \\
5  & 61.240 & 58.890 & 54.174 \\
6  & 68.97 &  56.74 &  37.15 \\
7  & 0.292 &   0.449 &   0.596 \\
8  & 5.2 &    4.5 &    3.6 \\
9  & 0.48 &    0.58 &    0.51 \\
10 & 6.8 &    7.5 &    10.4 \\
\hline\hline
\end{tabular}
\end{table}

\begin{figure}
\begin{center}
\includegraphics[width=\columnwidth]{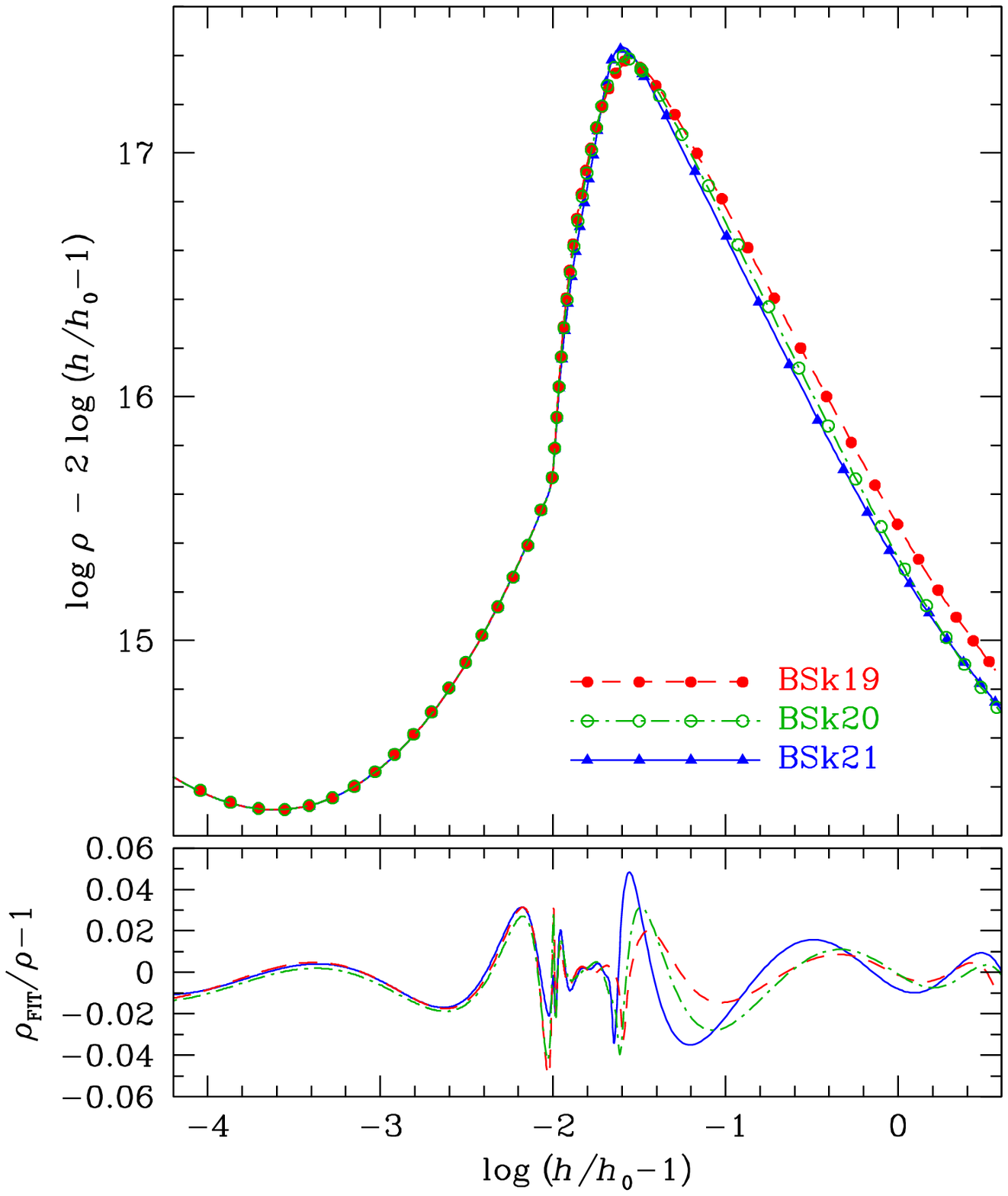}
\end{center}
\caption{Mass density as a function of pseudo-enthalpy.
Upper panel: rarefied data to be fitted, calculated by
integration according to Eqs.~(\ref{H.def}) and
(\ref{fit.P})
(symbols), compared with the fit (\ref{fit.rho-h}) (lines). 
Lower panel: fractional difference between the data and fit.
The symbols and line styles for BSk19, BSk20, and BSk21 are
the same as in Fig.~\ref{fig:Pfit}.
}
\label{fig:hfit}
\end{figure}

A comparison of the fit and the data is presented  in
Fig.~\ref{fig:hfit}. The typical fit error of $\rho$ is
about 1\% in the interval
$6\times10^{-5}\lesssim\eta\lesssim4$, which corresponds to
the considered mass density range
$10^6~\gcc\lesssim\rho\lesssim10^{16}$~\gcc. The maximum fit
errors of (3\,--\,5)\% occur, as expected, near physical
discontinuities, where the slope of the function $\xi(\eta)$
quickly changes: at  $\eta\sim0.01$, where the fit goes
smoothly through a break at the neutron-drip point, and at
$\eta\sim0.03$\,--\,0.1, near the crust-core interface.

%%%%%%%%%%%%%%%%%%%%%%%%%%%%%%%%%%%%%%%%%%%%%%%%%%%%%%%%%%%%%%%%%%%%%%%%
\section{Particle number fractions and density distributions}
\label{sect:frac}

The physical input for numerical simulations of neutron-star
thermal evolution includes the heat capacity, neutrino
emissivity, and thermal conductivity tensors in the crust
and the core \citep[see][for reviews]{YakPethick,PageGW}. 
The evolution of the magnetic field is coupled to
the thermal evolution and  depends on the electrical
conductivity tensor 
\citep[e.g.,][and references therein]{PonsMG}. 
For calculation of these quantities, it is important to know
the nucleon distributions in the crust and the composition of the core.

%%%%%%%%%%%%%%%%%%%%%%%%%%%%%%%%%%%%%%%%%%
\subsection{Nucleons and leptons in the core}
\label{sect:fracore}

%%%%%%%%%%%%%%%%%%%%%%%%%%%%%%%%%%%%%%%%%%%%%%%%%%%%%%%%%%%%%%%%%%%%%%%%
%%%%%%%%%%%%%%%%%%%%%%%%%%%%%%%%%%%%%%%%%%%%%%%%%%%%%%%%%%%%%%%%%%%%%%%%

%%%%%%%%%%%%%%%%%%%%%%%%%%%%%%%%%%%%%%%%%%%%%%%%%%%%%%%%%%%%%%%%%%%%%%%%
\begin{figure}
\begin{center}
\includegraphics[width=\columnwidth]{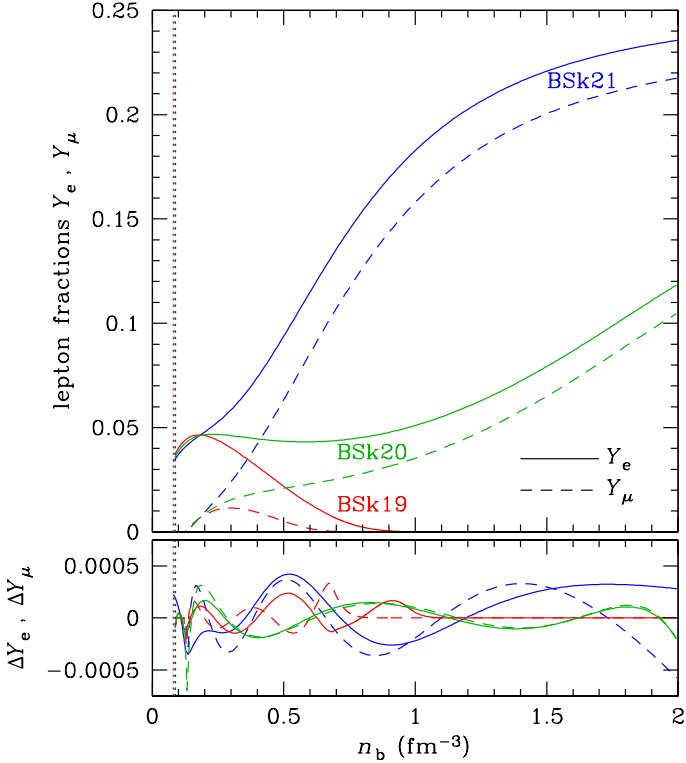}
\end{center}
\caption{Upper panel: Number fractions of electrons $Y_\mathrm{e}$
(solid lines) and muons $Y_\mathrm{\mu}$ (dashed lines) in the core
of a neutron star, relative to the number of nucleons, as
functions of nucleon number density $\nb$ for the three EoSs
indicated near the curves. Lower panel: Differences between
the fit (\ref{fit.Y})  and numerically computed tables for
$Y_\mathrm{e}$ and $Y_\mathrm{\mu}$, plotted against $\nb$.  The vertical
dotted lines show the position of the boundary between the
crust and the core ($\nc$ in Table~\ref{tab:fit.A}).
}
\label{fig:Yfit}
\end{figure}
%%%%%%%%%%%%%%%%%%%%%%%%%%%%%%%%%%%%%%%%%%%%%%%%%%%%%%%%%%%%%%%%%%%%%%%%

\begin{table}
\centering
\caption[]{Parameters of \req{fit.Y}.}
\label{tab:fit.Y}
\begin{tabular}{r|ccc}
\hline\hline\rule[-1.4ex]{0pt}{4.3ex}
$i$ &  BSk19 & BSk20 & BSk21  \\
\hline\rule[-1.4ex]{0pt}{3.7ex}
 & \multicolumn{3}{c}{$q_i^\mathrm{(e)}$}  \\
1  & $-0.0157$ & $-0.0078$ & 0.00575 \\
2  & 0.9063 & 0.745 & 0.4983  \\       
3  & 0 & 0.508 & 9.673  \\             
4  & 26.97 & 22.888 & 16.31 \\ % the last number is corrected 19.03.14
5  & 106.5 & 0.449 & 38.364  \\        
6  & 4.82 & 0.00323 & 0  \\            
\hline\rule[-1.4ex]{0pt}{3.7ex}
 & \multicolumn{3}{c}{$q_i^\mathrm{(\mu)}$} \\
1  & $-0.0315$ & $-0.0364$ & $-0.0365$ \\
2  &   0.25 & 0.2748 & 0.247 \\
3  &   0 & 0.2603 & 11.49 \\
4  &   12.42 & 12.99 & 24.55 \\
5  &   72.4 & 0.0767 & 48.544 \\
6  &   19.5 & 0.00413 & 0 \\
\hline\hline
\end{tabular}
\end{table}

The core of a neutron
star contains free neutrons, protons, electrons,
and muons, whose number fractions relative
to the total number of nucleons are, respectively, $1-Y_\mathrm{p}$,
$Y_\mathrm{p}$, $Y_\mathrm{e}$, and $Y_\mathrm{\mu}$. The condition of
electric neutrality requires that $Y_\mathrm{p} = Y_\mathrm{e}+Y_\mathrm{\mu}$. For $Y_\mathrm{e}$
and $Y_\mathrm{\mu}$, shown in Fig.~\ref{fig:Yfit},
we obtained the following fitting expression:
\begin{equation}
   Y_\mathrm{e,\mu} = \frac{
           q_1^\mathrm{(e,\mu)} + q_2^\mathrm{(e,\mu)} n + q_3^\mathrm{(e,\mu)}
                          n^4}{
          1+ q_4^\mathrm{(e,\mu)} n^{3/2} + q_5^\mathrm{(e,\mu)} n^4}
       \, \exp\left(-q_6^\mathrm{(e,\mu)} n^5 \right),
\label{fit.Y}
\end{equation}
where $n=\nb/\textrm{fm}^{-3}$, and parameters
$q_i^\mathrm{(e,\mu)}$ are given in Table~\ref{tab:fit.Y}. 
Whenever \req{fit.Y} gives a negative value, it must be
replaced by zero.  At the densities in the core, this can
occur for muons, when the muon chemical potential is smaller
than the electron rest energy. The fit residuals are
typically (1\,--\,3)$\times10^{-4}$ with a maximum of
$7\times10^{-4}$. Here, unlike in the preceding section, we
quote absolute (not fractional) errors, because the
fractional error loses its sense for a quantity that may be
zero, as $Y_\mathrm{\mu}$. The deviations of the fit from the data are
displayed in the bottom panel of Fig.~\ref{fig:Yfit}. 

As well known, fractional abundances of leptons in the
neutron-star core directly affect the thermal evolution of a
neutron star at the neutrino cooling stage (e.g.,
\citealp{YakPethick,PageGW}). In the region of a
neutron-star core where the Fermi momenta of protons
($p_\mathrm{Fp}$), neutrons ($p_\mathrm{Fn}$), and electrons
($p_\mathrm{Fe}$) satisfy the triangle inequality
$|p_\mathrm{Fp}-p_\mathrm{Fe}| < p_\mathrm{Fn} <
p_\mathrm{Fp}+p_\mathrm{Fe}$, the direct Urca (durca)
process of neutrino emission overpowers the more common
modified Urca process and greatly accelerates the cooling of
the star \citep{Lattimer-ea91,Haensel95}. As discussed by
\citet{Chamel-ea11}, the triangle inequality is never
satisfied in the BSk19 model, but the models BSk20 and BSk21
allow it at sufficiently high densities
$\nb>n_\mathrm{durca}$. The fitting expressions
(\ref{fit.Y}) allow us to reproduce the thresholds
$n_\mathrm{durca}=1.49$~fm$^{-3}$ for BSk20 and
$n_\mathrm{durca}=0.45$~fm$^{-3}$ for BSk21
\citep{Chamel-ea11} with  discrepancies of 0.08\% and 0.3\%,
respectively.

%%%%%%%%%%%%%%%%%%%%%%%%%%%%%%%%%%%%%%%%%%
\subsection{Nucleon numbers in the crust}
\label{sect:crust}

The outer crust consists of separate shells of 
different isotopes. The nuclear
mass
numbers $A$ and charge numbers $Z$ are constant
within each shell and differ from one shell to another.
The
nuclei are embedded in the sea of degenerate electrons.
Layers of diffusive mixing between adjacent shells are very
narrow (see \citealp{HameuryHB83}) and, therefore,
unimportant for most applications.
In particular, 
\citet{PearsonGC11} have presented tables of
$Z$ and $A$ in the outer crust for the models BSk19, BSk20,
and BSk21. 

In the case of the inner crust, the EoS of
\citet{Pearson-ea12} is based on the TETFSI
(temperature-dependent extended Thomas-Fermi Strutinsky
integral) method of \citet{Onsi-ea08}. This is a
semi-classical approximation to the
HFB method, with proton shell
corrections added perturbatively. A spherical Wigner-Seitz
cell is assumed, with the neutron and proton density
distributions parametrized according to
\begin{equation}
   n_q(r) = n_{0q}\,f_q(r) + n_{\mathrm{out},q},
\label{n_q}
\end{equation}
where $r$ is the distance to the center of the Wigner-Seitz
cell, $q={}$n for 
neutrons and $q={}$p for protons, $n_{\mathrm{out},q}$ represents
a constant background term, while $f_q(r)$ is a damped version of the usual
Fermi profile. Specifically, we write it as 
\begin{equation}
   f_q(r) = \left[1+D_q(r)\,\exp\left(\frac{r-C_q}{a_q}\right) \right]^{-1},
\label{nucform}
\end{equation}
where $C_q$ is the width of $f_q(r)$ at half maximum and $a_q$ is the
diffuseness parameter, while $D_q(r)$ is a damping factor given by
\begin{equation}
\label{damp}
D_q(r) = \exp\left\{\left(\frac{C_q-\RWS}{r-\RWS}\right)^2
-1 \right\} ,
\end{equation}
where  $\RWS$ is the Wigner-Seitz cell radius The purpose of
this factor is to ensure that $f_q(r)$ and all its
derivatives will vanish at the surface of the cell, a
necessary condition for the validity of the semi-classical
approximations that underlie the TETFSI method. 

But whether we include this damping factor, or set it equal to unity, leaving
us with a simple Fermi profile, the parametrization of Eq.~(\ref{n_q})
eliminates an arbitrary separation into liquid and gaseous phases within the 
Wigner-Seitz cell, and, strictly speaking, makes it illegitimate to draw a
distinction between a ``neutron gas" and ``nuclei". Nevertheless, if we denote
by $N^\prime$ the total number of neutrons in the cell at a given density, it
is convenient to define the number of cluster neutrons as
\begin{equation}
N = N^\prime - n_{\mathrm{out,n}}V_\mathrm{WS}, 
\end{equation}
where $V_\mathrm{WS}=4\pi\RWS^3/3$ is the volume of the
Wigner-Seitz cell, with a similar  expression for the number
of cluster protons $Z$. Actually, almost everywhere in the
inner crust $n_{\mathrm{out,p}} = 0$, so that $Z =
Z^\prime$, the total number of protons in the cell. However,
near the transition to the core some  protons tend to spread
over the entire cell, so that $Z<Z'$. This spreading
corresponds to a smooth (second or higher order) phase
transition between the  crust and the core suggested by
\citet{Pearson-ea12}.
 
It turns out that for all three models considered here the number of protons 
$Z^\prime$ in the Wigner-Seitz cell is equal to 40 for all densities. However, 
the number of neutrons $N^\prime$ in the cell varies considerably with density,
and, in fact, will be nonintegral in general since it is taken in the TETFSI
method as one of the variables with respect to which the
energy is minimized.
The notion of a fractional number of neutrons per cell corresponds to
the  physical reality of delocalized neutrons. On the other hand, since the 
TETFSI method calculates proton (but not neutron) shell effects, $Z^\prime$
cannot be allowed to become nonintegral, even when the protons become 
delocalized.

With $N$, $Z$, and $A^\prime = N^\prime + Z^\prime$ varying smoothly with
$n_b$,  we obtained the following parametrizations over the inner crust (note
that $Z^\prime$ = 40 everywhere):
\begin{eqnarray}
   A-Z &=&
 \frac{
        p_1 + p_2 \log x + (p_3 \log x)^{3.5}
         }{
           1 + (p_4 \log x)^{p_5}}
                   \big[1-(\nb/\nc)^{p_0}\big],
\label{fit.A}
\\
  Z &=& Z'\,\exp\left( -(\nb/\nc)^{p''_1} \right)
                   \big[1-(\nb/\nc)^{p''_0}\big],
\label{fit.Z}
\\
   A' &=& \frac{
        p'_1 + p'_2 \log x +p'_3 (\log x)^2
         }{ 1 + (p'_4 \log x)^4 }
            (1 + p'_5 x) \left[1-(p'_6 x)^2\right],
\label{fit.A'}
\end{eqnarray}
Here, $x = \nb/\nd$, $\nd$ is the baryon number density at the neutron-drip 
point, and $\nc$ is the density of the transition to the homogeneous core. The 
parameters $p_i$, $p'_i$, $p''_i$ are listed in Table~\ref{tab:fit.A} along 
with the densities $\nd$ and $\nc$. The square brackets with the large power 
indices $p_0$ in \req{fit.A} and $p_0''$ in \req{fit.Z} ensure the quick
decrease of $A$ and $Z$ to zero in a narrow density interval at the transition 
from the crust to the core; they almost do not affect the fits in the bulk of 
the crust.
           
\begin{table}
\centering
\caption[]{Fitting
parameters for Eqs.~(\ref{fit.A}),  (\ref{fit.Z}), and
(\ref{fit.A'}). The last two lines list
the number densities of baryons
at the neutron drip point,
$\nd$, and
at the crust/core boundary, $\nc$ (in fm$^{-3}$)
\citep{Pearson-ea12}.}
\label{tab:fit.A}
\begin{tabular}{r|ccc}
\hline\hline\rule[-1.4ex]{0pt}{4.3ex}
  & BSk19 & BSk20 & BSk21  \\
\hline\rule[-1.4ex]{0pt}{3.7ex}
$i$ & \multicolumn{3}{c}{$p_i$}\\
0 & 8.40 & 9.30 & 10.8 \\
1 & 93.0 & 92.8 & 92.3 \\
2 & 11.90 & 12.95 & 13.80 \\
3 & 1.490 & 1.493 & 1.625 \\
4 & 0.334 & 0.354 & 0.3874 \\
5 & 5.05 & 7.57 & 13.8 \\
\hline\rule[-1.4ex]{0pt}{3.7ex}
$i$ & \multicolumn{3}{c}{$p'_i$}\\
1 & 134.7 & 134.7 & 132.6 \\
2 & 183.7 & 188.2 & 187.6 \\
3 & 308.7 & 275.6 & 229.2 \\
4 & 0.3814 & 0.4346 & 0.5202 \\
5 & $-5.8\times10^{-4}$ & 0.00163 & 0.00637 \\
6 & $4.9\times10^{-4}$ & 0.00149 & 0.00151 \\
\hline\rule[-1.4ex]{0pt}{3.7ex}
$i$ & \multicolumn{3}{c}{$p''_i$}\\
0 & 16 & 20 & 27 \\
1 & 19 & 19 & 17 \\
\hline\rule[-1.4ex]{0pt}{3.7ex}
$\nd$ & $2.63464\times10^{-4}$ &
$2.62873\times10^{-4}$ & $2.57541\times10^{-4}$ \\
$\nc$ & 0.0885 & 0.0854 & 0.0809 \\
\hline\hline
\end{tabular}
\end{table}

%%%%%%%%%%%%%%%%%%%%%%%%%%%%%%%%%%%%%%%%%%%%%%%%%%%%%%%%%%%%%%%%%%%%%%%%
\begin{figure}
\begin{center}
\includegraphics[width=\columnwidth]{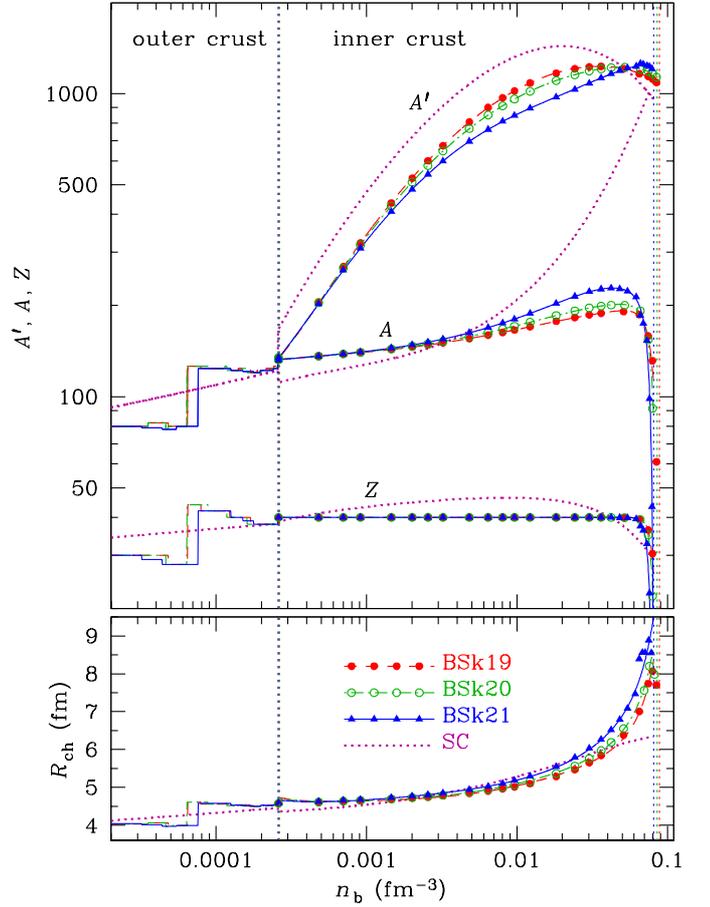}
\end{center}
\caption{Effective nuclear parameters in the neutron-star
crust, according to different models, as
functions of baryon number density $\nb$ in
the crust of a neutron star.
Top panel: nuclear mass number
$A$ (the middle group of curves), the number of protons in a
cluster $Z$ (the lower curves), 
and the effective Wigner-Seitz cell mass number
$A'$ (the upper curves).
Bottom panel: rms charge radius of the nucleus.
The results are shown for the models
BSk19 (dashed lines and filled dots),
BSk20 (dot-dashed lines and open circles), BSk21 (solid
lines and filled triangles), and, for comparison,
the smooth-composition model (the dotted curves).
 The left vertical dotted line shows the position
of the boundary between the inner and outer crust,
$\nb=\nd$ (to the left of this line $A'=A$). The right vertical
dotted lines mark $\nc$, the interface between the
crust and the core, in the
BSk models. Between these boundaries, the
curves represent analytical fits, and the symbols show
some of the numerical data.
}
\label{fig:bsksc}
\end{figure}
%%%%%%%%%%%%%%%%%%%%%%%%%%%%%%%%%%%%%%%%%%%%%%%%%%%%%%%%%%%%%%%%%%%%%%%%

In the upper panel of Fig.~\ref{fig:bsksc} we show the
numbers $A'$, $A$, and $Z$ as functions of $\nb$ in the outer
and inner crust of a neutron star. In the lower panel we
show the rms radius of the charge
distribution, $\Rc$. The values obtained in the
models BSk19, BSk20, and BSk21 are compared with the
smooth-composition (SC) model (Appendix~B.2 of
\citealt{HPY}; see the next section). The vertical dotted
lines mark the position of the neutron-drip point, which
separates the inner and outer crust, and the transition
from the crust to the core. 
In the inner crust, the fitted values (lines) are compared
with the numerical results (symbols) for the BSk models.
The numerical data are rarefied, in order to
avoid crowding of the symbols. The agreement between the
data and the fits are typically within a fraction of
percent, except for a vicinity of the boundary with the
core.  Near the latter boundary, at
0.065~fm$^{-3}\lesssim\nb<\nc$, the neutron and proton
distributions become rather flat (see Fig.~\ref{fig:nucform}
below), which hampers an accurate determination of the shape
parameters. Therefore the numerical data suffer a
significant scatter in this density interval, but the fits
show an acceptable qualitative agreement with them. In the
outer crust (at $\nb<\nd$) we use the elemental composition
$A$ and $Z$ from \citet{PearsonGC11} and $\Rc$
from \textsc{bruslib}.\footnote{\raggedright{The Brussels
Nuclear Library for Astrophysics Application,
http://www.astro.ulb.ac.be/bruslib/}}
{}From Fig.~\ref{fig:bsksc} we see that the SC model predicts a
considerable jump of $A'$ at the neutron-drip point,
while in the BSk models $A'$ is almost
continuous. The latter continuity is equivalent to the
equality of $\RWS$ at both sides of the drip boundary.

%%%%%%%%%%%%%%%%%%%%%%%%%%%%%%%%%%%%%%%%%%%%%%%%%%%%%%%%%%%%%%%%%%%%%%%%
\subsection{Nuclear shapes and sizes in the crust}
\label{sect:shape}

In applications one sometimes needs a more detailed
information about microscopic distributions of nucleons than
given by the numbers $A$, $A'$, $Z$, and $Z'$ considered in
Sect.~\ref{sect:crust}. For example, cross sections of
scattering of charged particles depend on the charge
distribution in a nucleus. Previous calculations of neutrino
bremsstrahlung \citep[e.g.,][]{Kam-ea99} and electron heat
conduction \citep[e.g.,][]{GYP} in the crust of a neutron
star employed nuclear form factors provided by the SC model
based on the parametrization~(\ref{n_q}) with   $f_q(r) = [1
- (r/r_{\mathrm{max},q})^{b_q}]^3$~\citep{Oyamatsu93}. Note
that  $n_{\mathrm{out,p}}$ is effectively zero, and
$n_{\mathrm{out,n}} = (A'-A)/V_\mathrm{WS}$.
\citet{Kam-ea99} fitted $\RWS$, $n_{0q}$,
$r_{\mathrm{max},q}$, and $b_q$ as functions of $\nb$ by
interpolating the \citet{Oyamatsu93} values of these
parameters through the inner crust and making use of the
\citet{HP94} data for the outer crust.

%%%%%%%%%%%%%%%%%%%%%%%%%%%%%%%%%%%%%%%%%%%%%%%%%%%%%%%%%%%%%%%%%%%%%%%%
\begin{figure}
\begin{center}
\includegraphics[width=\columnwidth]{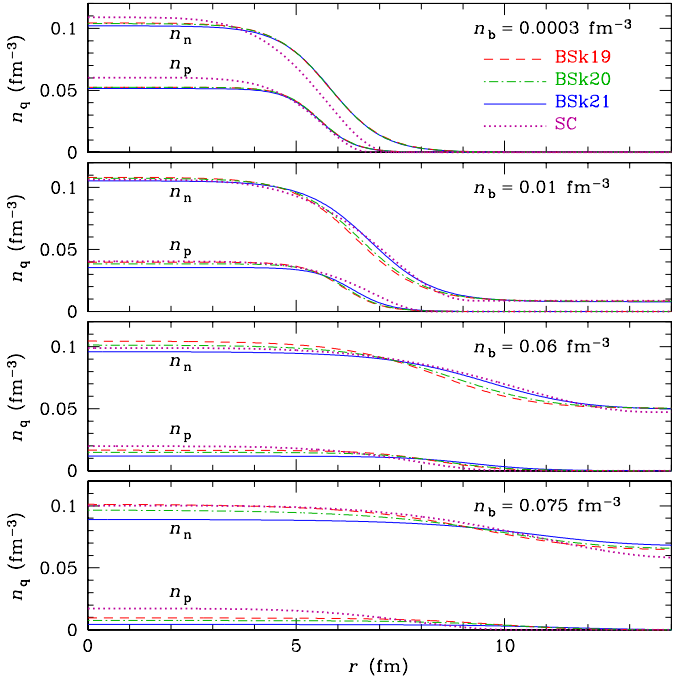}
\end{center}
\caption{Neutron ($n_\mathrm{n}$) and proton ($n_\mathrm{p}$) density profiles
at the values of the average baryon density 
$\nb=3\times10^{-4}$ fm$^{-3}$ (near the top of the inner
crust), 0.01 fm$^{-3}$, 0.06 fm$^{-3}$, and 0.075 fm$^{-3}$
(near the bottom of the inner crust), according to the
models BSk19, BSk20, and BSk21. The SC model
\citep{HPY} is also shown for comparison.
}
\label{fig:nucform}
\end{figure}
%%%%%%%%%%%%%%%%%%%%%%%%%%%%%%%%%%%%%%%%%%%%%%%%%%%%%%%%%%%%%%%%%%%%%%%%

In Fig.~\ref{fig:nucform} we show neutron and proton density
distributions  near the center of a Wigner-Seitz cell as
given by \citet{Pearson-ea12}, i.e., as described by
Eqs.~(\ref{nucform}) and  (\ref{damp}) for the models
BSk19,  BSk20, and BSk21. The four panels correspond to
four different values of the mean baryon density $\nb$ in
the inner crust. The neutron and proton density  profiles
predicted by the SC model are also shown for comparison. We
present  the parameters $C_q$, $a_q$, and $n_{0q}$ that
enter Eqs.~(\ref{n_q}) and  (\ref{nucform}) as functions of
$n=\nb/\textrm{fm}^{-3}$,
\begin{equation}
  y
            =
       \frac{s_1+s_2\,n^{s_3}}{
         1-s_4\,n^3},
\label{fitshape}
\end{equation}
where $y=C_\mathrm{p}$, $a_\mathrm{p}$, $C_\mathrm{n}$, or $a_\mathrm{n}$, with numerical parameters $s_i$ listed 
in Table~\ref{tab:fitshape}. The decreasing denominator ensures the
accelerated increase of the size and diffuseness of the nuclei when the density
approaches that of the uniform nuclear matter near the crust-core interface.
The values of $n_{0q}$ are related to the other fitted 
parameters via
\begin{equation}
   4\pi\,n_{0q}\,\int_0^{\RWS}\! f_q(r)\,r^2\,\dd r 
            = \left\{ \begin{array}{cl}
                A-Z & \mbox{for~} q=n, \\
                 Z  & \mbox{for~} q=p.
         \end{array} \right.
\end{equation}
Nevertheless we find it convenient for applications to have
separate simple fits
\begin{eqnarray}
   n_{0p} &=& \left(
         q^\mathrm{(p)}_1 - \frac{q^\mathrm{(p)}_2 \, n }{ q^\mathrm{(p)}_3 + n^{q^\mathrm{(p)}_4}}
             \right)\,
           \big[  1-(\nb/\nc)^9  \big],
\label{fitn0p}
\\
   n_{0n} &=&  \left(
        q^\mathrm{(n)}_1 + q^\mathrm{(n)}_2 \sqrt{n} - q^\mathrm{(n)}_3\,n
             \right)\,
           \big[  1-(\nb/\nc)^{16}  \big],
\label{fitn0n}
\end{eqnarray}
with parameters $q^\mathrm{(p)}_i$ and $q^\mathrm{(n)}_i$ listed in
Table~\ref{tab:fitn0q}.
The agreement between the
data and the fits are typically $\sim1$\% or better, 
except for a bottom crust layer,
as explained in the previous section.

\begin{table}
\centering
\caption[]{Parameters $s_i$ of \req{fitshape} for $C_q$ and $a_q$.}
\label{tab:fitshape}
\begin{tabular}{r|ccc|ccc}
\hline\hline\rule[-1.4ex]{0pt}{4.3ex}
$i$ &  BSk19 & BSk20 & BSk21 &  BSk19 & BSk20 & BSk21  \\
\hline\rule[-1.4ex]{0pt}{3.7ex}
 & \multicolumn{3}{c|}{$s_i(C_\mathrm{p})$}  & \multicolumn{3}{c}{$s_i(a_\mathrm{p})$} \\
1  & 5.500 & 5.493 & 5.457    & 0.4377 & 0.4353 & 0.4316 \\
2  & 11.7 & 12.8 & 14.2       &  4.360 & 4.440 & 4.704 \\
3  & 0.643 & 0.636 & 0.601    &  1 & 1 & 1 \\
4  & 472  & 484   & 566       & 1084 & 1154 & 1253 \\
\hline\rule[-1.4ex]{0pt}{3.7ex}
 & \multicolumn{3}{c|}{$s_i(C_\mathrm{n})$}  &
\multicolumn{3}{c}{$s_i(a_\mathrm{n})$} \\
1  & 5.714 & 5.714 & 5.728  & 0.639 & 0.632 & 0.636 \\
2  &14.05 & 16.3 & 22.2     & 1.461 & 1.98  & 5.32 \\
3  & 0.642 & 0.645 & 0.663  & 0.457 & 0.514 & 0.739 \\
4  & 182 & 175 & 144        & 1137 & 1122 & 624 \\
\hline\hline
\end{tabular}
\end{table}

\begin{table}
\centering
\caption[]{Parameters of Eqs.~(\ref{fitn0p}) and
(\ref{fitn0n}).}
\label{tab:fitn0q}
\setlength{\tabcolsep}{5pt}
\begin{tabular}{r|ccc|ccc}
\hline\hline\rule[-1.4ex]{0pt}{3.7ex}
 & \multicolumn{3}{c|}{$q^\mathrm{(p)}_i$}  & \multicolumn{3}{c}{$q^\mathrm{(n)}_i$} \\
$i$ &  BSk19 & BSk20 & BSk21 &  BSk19 & BSk20 & BSk21  \\
\hline\rule{0pt}{2.3ex}
1  & 0.05509 & 0.05382 & 0.05273  & 0.10336 & 0.10283 & 0.10085 \\
2  & 0.1589 & 0.1400 & 0.1107 & 0.0772 & 0.0825 & 0.0942 \\
3  &    0 & 0.01715  & 0.02218  & 1.129 & 1.189 & 1.279 \\
4  & 0.4917  & 0.566  & 0.6872   & -- & -- & -- \\
\hline\hline
\end{tabular}
\end{table}

\begin{table}
\centering
\caption[]{Parameters $s_i$ of \req{fitshape} for $x_\mathrm{nuc}$ 
and $x_{\mathrm{nuc,n}}$.}
\label{tab:fitx}
\begin{tabular}{r|ccc|ccc}
\hline\hline\rule[-1.4ex]{0pt}{3.7ex}
 & \multicolumn{3}{c|}{$s_i(x_{\mathrm{nuc}})$}
  & \multicolumn{3}{c}{$s_i(x_{\mathrm{nuc,n}})$} \\
$i$ &  BSk19 & BSk20 & BSk21 &  BSk19 & BSk20 & BSk21  \\
\hline\rule{0pt}{2.2ex}
1  & 0.1120 & 0.1094 & 0.1045 & 0.122 & 0.119 & 0.114 \\
2  & 2.06 & 2.04 & 2.09 & 2.27 & 2.30 & 2.56 \\
3  & 0.633 & 0.613 & 0.586 & 0.618 & 0.603 & 0.595 \\
4  & 507 & 509 & 513 & 193 & 182 & 107 \\
\hline\hline
\end{tabular}
\end{table}

%%%%%%%%%%%%%%%%%%%%%%%%%%%%%%%%%%%%%%%%%%
\section{Conductivity in the neutron-star crust}
\label{sect:cond}

As an example of application of our fitting formulae, we
consider the electron conductivities in the crust of a
neutron star. 
Practical formulae for calculation of the electron
electrical and thermal conductivities of fully ionized
plasmas have
been developed by \citet{Potekhin-ea99} in the approximation
of pointlike nuclei (see references therein for earlier
works) and extended by \citet{GYP} to the case where the
finite nuclear size cannot be neglected. The latter authors
considered different nuclear shape models and concluded that
their effect on electrical conductivity $\sigma$ is mainly governed
by the ratio $x_\mathrm{nuc}=R_\mathrm{eff}/\RWS$, where
$R_\mathrm{eff}=\sqrt{5/3}\, \Rc$ is the effective
radius of the uniformly charged sphere that has the same
rms radius of the charge
distribution, $\Rc$, as the considered nucleus.
The same parameter $x_\mathrm{nuc}$
determines the effect of nuclear form factors on
the thermal conductivity $\kappa$, except at low
temperatures where quantum lattice effects violate the
Wiedemann-Franz relation in a nontrivial way \citep{GYP}. 
Under the assumption that the charge distribution is proportional to 
the proton distribution,
\begin{equation}
    x_\mathrm{nuc}^2 = \frac53\,\frac{
          \int_0^{\RWS}\!f_\mathrm{p}(r)\,r^4\,\dd r }{
           \RWS^2 \int_0^{\RWS}\!f_\mathrm{p}(r)\,r^2\,\dd r },
\label{xnuc}
\end{equation}
we obtained a direct fit to this quantity, which
facilitates calculation of the conductivities. In addition,
we calculated and fitted the analogous size parameter
$x_{\mathrm{nuc,n}}=R_{\mathrm{eff,n}}/\RWS$ for  the
neutron distribution $f_\mathrm{n}(r)$. The fitting was done
for the quantities obtained from
Eqs.~(\ref{xnuc}) and (\ref{nucform}) with calculated (not fitted) values of $C_q$,
$a_q$, and $\RWS$ at each density. The fits have the form of
\req{fitshape} with the parameters listed in
Table~\ref{tab:fitx}. The
residuals are similar, with maxima of several percent near the end
of the $\nb$ range and an order of magnitude smaller typical
values.

%%%%%%%%%%%%%%%%%%%%%%%%%%%%%%%%%%%%%%%%%%%%%%%%%%%%%%%%%%%%%%%%%%%%%%%%
\begin{figure}
\begin{center}
\includegraphics[width=\columnwidth]{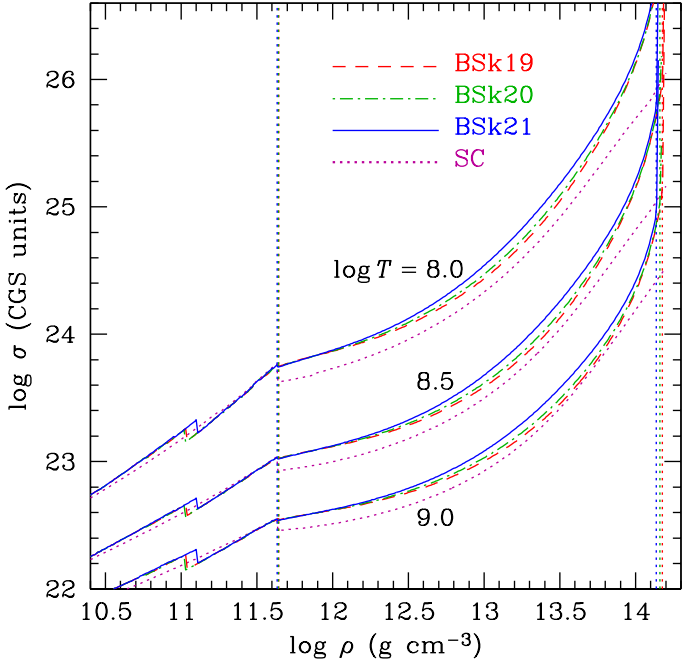}
\end{center}
\caption{Electrical conductivity $\sigma$ of the
neutron-star crust with nuclear parameters given by the
models BSk19 (dashed lines), BSk20 (dot-dashed lines), BSk21
(solid lines), and the smooth-composition (SC) model
(dotted curves) as functions of mass density
$\rho$ for $T=10^8$~K, $10^{8.5}$~K, and $10^9$~K. The
vertical dotted lines show the boundaries $\nd$ and $\nc$ of
the inner crust according to the three BSk models
(Table~\ref{tab:fit.A}).
}
\label{fig:cond}
\end{figure}
%%%%%%%%%%%%%%%%%%%%%%%%%%%%%%%%%%%%%%%%%%%%%%%%%%%%%%%%%%%%%%%%%%%%%%%%

Figure~\ref{fig:cond} displays electrical conductivities
$\sigma$ calculated with the fitted $A'$, $A$, $Z$, and
$x_\mathrm{nuc}$ at temperatures and densities
characteristic of the outer and inner crusts of neutron
stars. Here, the calculations are performed for a
body-centered cubic lattice of the nuclei without
impurities. The crystalline structure of the crust is
favored by recent results of molecular-dynamics simulations
\citep{Hughto-ea11} and supported by comparison of 
observations of X-ray transients in quiescence with
simulations of the cooling of their crust
\citep{Shternin-ea07,Cackett-ea10}. We have removed
switching from Umklapp to normal processes of
electron-phonon scattering at low $T$ from the previously
developed code \citep{GYP}, since  \citet{Chugunov12} has
shown that the normal processes have no effect under the
conditions in a neutron-star crust. Figure~\ref{fig:cond}
shows that the BSk models predict typically 1.5\,--\,2 times
higher conductivity $\sigma$ in the inner crust, than the SC
model. On the other hand, this
difference in $\sigma$ can be removed by allowing a
modest impurity parameter  $Z_\mathrm{imp} = \langle
(Z-\langle Z\rangle)^2 \rangle^{1/2} \sim 3$. The
difference between the BSk and SC models increases
near the bottom of the crust, where the
BSk models (unlike SC) provide a continuous transition to the
uniform nuclear matter in the stellar core.
The conductivity is also almost
continuous at the neutron drip density in the BSk models, at
contrast to the SC model where it abruptly
decreases at the drip point. 

Similarly, we have used the analytical fits for calculating thermal
conductivity $\kappa$. It has to be noted that electron
conduction is probably the main mechanism of heat
transport in the entire neutron-star crust despite the
existence of competing transport channels. Radiative conduction is
negligible at the crust densities
\citep{PotekhinYakovlev01,Perez-ea06}, the heat transport by
phonons is generally weaker than the electron transport 
\citep{Perez-ea06,ChugunovHaensel07}, and the heat transport
by superfluid neutrons in the inner crust 
\citep{BisnoRomanova,Aguilera-ea09} may be suppressed because
of the strong coupling of the neutron superfluid to the
nuclei due to nondissipative entrainment effects
\citep{Chamel12}. We have found that the difference between
the BSk and SC models for the electron heat conductivity
$\kappa$ is smaller than the corresponding difference for
$\sigma$. The smaller difference is explained by the large contribution of
 the electron-electron scattering into the thermal
resistivity of the inner crust \citep{ShterninYakovlev06}. 

%%%%%%%%%%%%%%%%%%%%%%%%%%%%%%%%%%%%%%%%%%
\section{Neutron star configurations}
\label{sec:fit_ns}

%%%%%%%%%%%%%%%%%%%%%%%%%%%%%%%%%%%%%%%%%%%%%%%%%%%%%%%%%%%%%%%%%%%%%%%%
\begin{figure*}
\centering
\includegraphics[width=.8\textwidth]{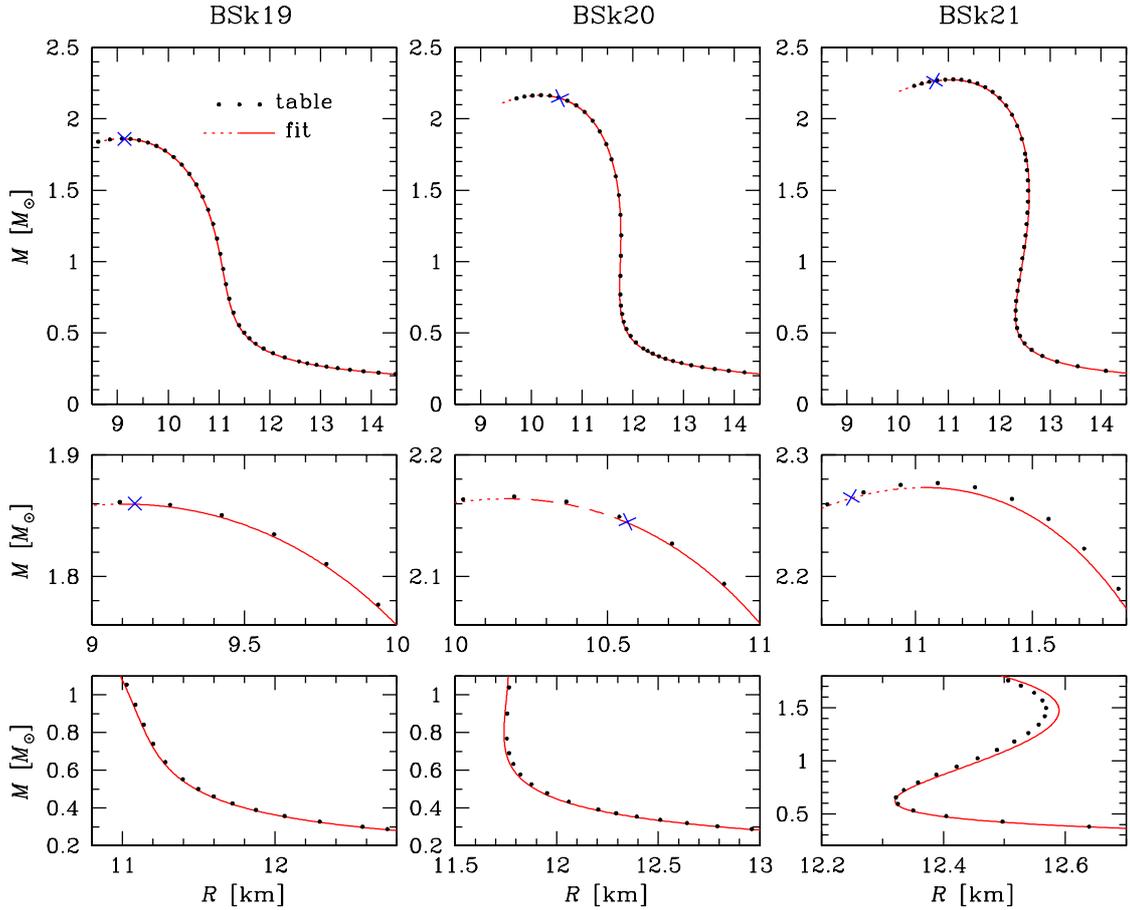}
\caption{Top panels: gravitational mass (in solar masses)
versus circumferential radius of nonrotating
neutron stars  for the EoSs BSk19-20-21 (dots) and their
analytical representations from Eq.~(\ref{fit.P}) (lines). 
The solid and dotted parts of the lines correspond
to the hydrostatically stable and unstable configurations,
respectively. The dashed segment in the middle panel
corresponds to superluminal EoS at the stellar center.
The crosses mark the threshold beyond which the EoS becomes
superluminal. The middle and bottom panels show respectively a
zoom around the maximum neutron-star mass and the low mass
region where the  discrepancies are the largest.}
\label{fig:MR_fitnonrot}
\end{figure*}
%%%%%%%%%%%%%%%%%%%%%%%%%%%%%%%%%%%%%%%%%%%%%%%%%%%%%%%%%%%%%%%%%%%%%%%%

%%%%%%%%%%%%%%%%%%%%%%%%%%%%%%%%%%%%%%%%%%%%%%%%%%%%%%%%%%%%%%%%%%%%%%%%
\begin{figure*}
\centering
\includegraphics[width=.8\textwidth]{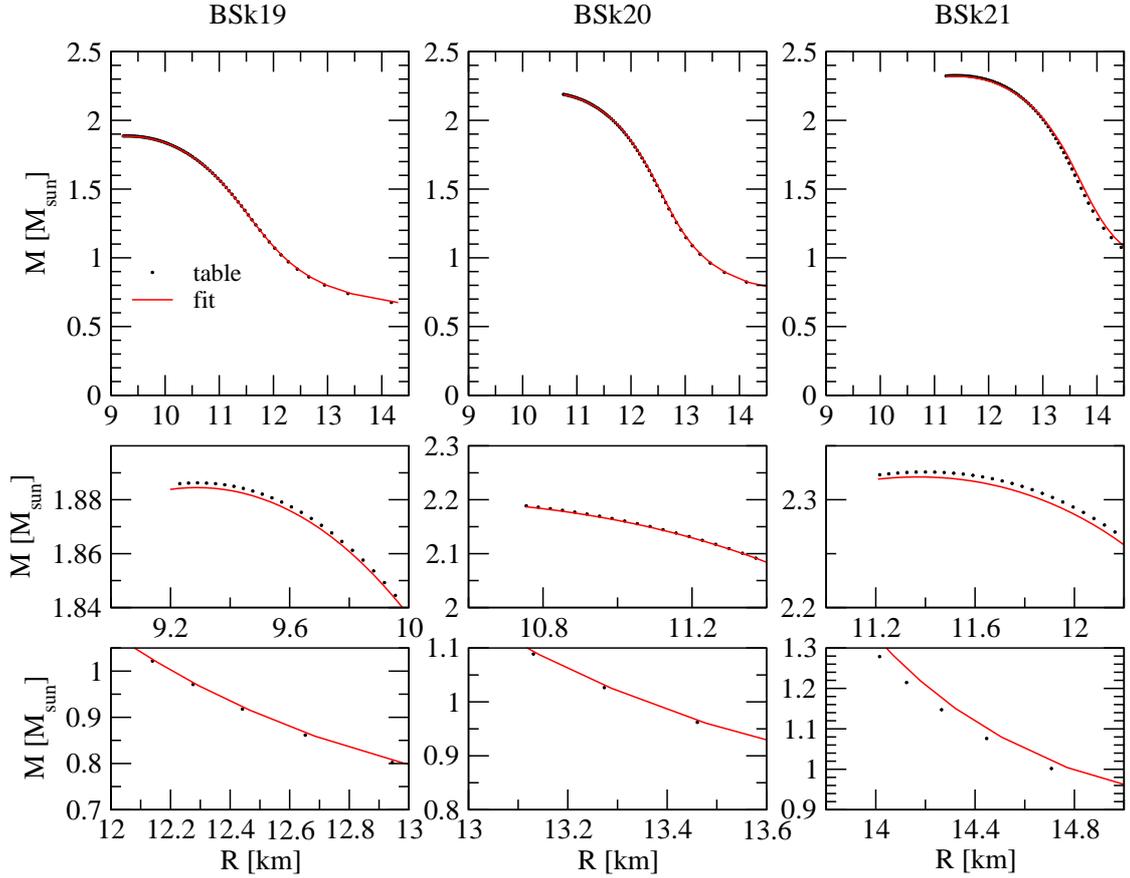}
\caption{Top panels: gravitational mass (in solar masses)
versus circumferential radius of rotating neutron stars 
(rotation frequency $716$~Hz) for the EoSs BSk19-20-21 (dots) and their analytical representations from Eq.~(\ref{fit.rho-h}) 
(solid lines). 
The middle and bottom panels show respectively a zoom around the maximum neutron-star mass and the low mass region where the 
discrepancies are the largest.}
\label{fig:MR_fitrot}
\end{figure*}
%%%%%%%%%%%%%%%%%%%%%%%%%%%%%%%%%%%%%%%%%%%%%%%%%%%%%%%%%%%%%%%%%%%%%%%%

In order to estimate the errors introduced in the fitting 
formulae described in Sect.~\ref{sect:EoS} on global
neutron-star  properties, we have computed the mass and 
radius of a neutron star from the tabulated EoSs and their 
analytical representations. For a nonrotating neutron star,
we integrated the Tolman-Oppenheimer-Volkoff (TOV) equation 
from the center, with the central mass density $\rhoc$
as a free parameter, outward to $\rho_\mathrm{s}$, using the
fourth-order Runge-Kutta method with an adaptive step and
controlled accuracy. We avoided an affixment of an
integration step to the calculated points using either
linear or cubic spline interpolation in the tabulated EoSs.
The neutron-star masses $M$ obtained using these two types
of interpolation agree to $\sim10^{-4}\,M_\odot$. The radii
$R$ agree typically to $\sim0.1$\%, if
$M\gtrsim0.3\,M_\odot$.

Figure~\ref{fig:MR_fitnonrot} shows the mass-radius relation
of nonrotating neutron  stars for the EoSs BSk19, BSk20, and
BSk21. The neutron-star configurations obtained  with the
original EoSs and with their analytical representations are
drawn as dots and lines respectively. The maximum
neutron-star masses are $M_\mathrm{max}=1.86\,M_\odot$ at
$\rhoc=3.48\times10^{15}$ \gcc{} for BSk19,
$M_\mathrm{max}=2.16\,M_\odot$ at $\rhoc=2.69\times10^{15}$
\gcc{} for BSk20,  and $M_\mathrm{max}=2.27\,M_\odot$ at
$\rhoc=2.27\times10^{15}$ \gcc{} for BSk21, in close
agreement with \citet{Chamel-ea11} and \citet{Fantina-ea12}.
The differences between $M_\mathrm{max}$ obtained using the
original data and the fit are about 0.09\%, 0.09\%, and
0.17\%, and the differences for corresponding $\rhoc$ values are about
0.03\%, 0.02\%, and 0.7\%, respectively. At higher $\rhoc$
the condition of hydrostatic stability $\dd M / \dd \rhoc$
is violated; the unstable configurations are shown by the
dotted parts of the curves to the left of the maxima in
Fig.~\ref{fig:MR_fitnonrot}. The crosses in
Fig.~\ref{fig:MR_fitnonrot} correspond to the largest values
of $\rhoc$ for which $\dd P/\dd\rho < c^2$
($2.18\times10^{15}$ \gcc{} for BSk19 and BSk20, and
$2.73\times10^{15}$ \gcc{} for BSk21). At higher densities
the EoS becomes superluminal. The fit (\ref{fit.P}) and its
first derivative determine these densities with accuracies
within 2\%. The corresponding stellar masses determined from
the fit and from the tables agree to $\sim0.1$\%.  Note that
the EoS BSk20 becomes superluminal before the limit of
hydrostatic stability is reached. Configurations with higher
$\rhoc$, corresponding to $2.14\,M_\odot<M<M_\mathrm{max}$
and shown by the dashed curve in
Fig.~\ref{fig:MR_fitnonrot}, cannot be fully trusted (see,
e.g.,  \citealp{HPY}, \S\,5.15, for a discussion)

The minimum neutron-star masses are $0.093\,M_\odot$,
$0.090\,M_\odot$, and $0.087\,M_\odot$ for the models BSk19,
BSk20, and BSk21, with discrepancies between the original
data and the fit $\sim0.7$\%, 0.1\%, and 0.03\%,
respectively. The radii of neutron stars with the mass of
$1.4\,M_\odot$ are $R=10.74$ km, 11.74 km, and 12.57 km,
with discrepancies between the original data and the fit of
$\sim0.1$\%, $\sim0.02$\%, and $\sim0.2$\%.
The discrepancies in the  circumferential radii ($R=11.5$
km, 11.9 km, and 12.4 km) of neutron stars with
$M=0.5\,M_\odot$ are also smaller than 0.2\% for all three EoSs.

As mentioned in \S\,\ref{sect:fracore} the triangle
condition for the durca processes can be fulfilled for BSk20
and BSk21 models at $\nb>n_\mathrm{durca}$. The threshold
$n_\mathrm{durca}$ cannot be reached in a stable neutron
star in frames of the model BSk20, but it is reached for
neutron stars with $M>1.59\,M_\odot$ in the model BSk21. The
latter mass value, first obtained by \citet{Chamel-ea11}, is
reproduced by the present fit with a discrepancy of 0.3\%.
\citet{Chamel-ea11} noted that all three EoSs are compatible
with the constraint that no durca process should occur in
neutron stars with masses 1\,--\,1.5\,$M_\odot$
\citep{Klaehn-ea06}. On the other hand,  according to the analysis
of \citet{Yak-ea08}, the situation where the most massive
neutron stars with nucleon superfluidity in the core
experience enhanced cooling due to the durca processes
appears in a better agreement with observations than the
complete absence of such processes in any stars. Thus the
model BSk21 may bring the cooling theory in a better
agreement with observations than the other models. The
fitting formulae presented above facilitate the checks of
this kind.
In this respect, it is worth noting that the effective
nucleon masses $m^*_\mathrm{n}$ and $m^*_\mathrm{p}$, which are needed for
cooling simulations, are readily obtained in analytical form
from Eq.~(A10) of \citet{ChamelGP09}, using the appropriate
parameter set given in \citet{Goriely-ea10}.

Similarly, we have analyzed the structure of a neutron star
rotating at a frequency of 716~Hz, equal to the frequency of
PSR J1748$-$2446ad, the fastest-spinning pulsar known
\citep{Hessels-ea06}. For this purpose we used the
\texttt{Lorene/Codes/Nrotstar/nrotstar} code from the public
library \textsc{lorene}.\footnote{\raggedright{Langage Objet pour la
Relativit\'e Num\'erique, http://www.lorene.obspm.fr/}}
We have generated tables of the analytical EoSs from
Eq.~(\ref{fit.rho-h}) and used them as an input in the
\texttt{nrotstar} code. The results obtained using the
original EoSs and their analytical representations are shown
in Fig.~\ref{fig:MR_fitrot}. Here we plot only the stable
stellar configurations that are described by subluminal
EoSs. The relative differences in the
maximum neutron-star masses are of similar magnitudes to
those found for static neutron stars, namely $\sim 0.03$\%, 
$\sim 0.08$\% and $\sim 0.2$\% for the EoSs BSk19, BSk20,
and BSk21. The errors in the radii of  $1\ M_\odot$ neutron
stars are $\sim 0.1$\%, $\sim 0.09$\%, and $\sim 0.5$\%,
respectively. 

All in all, the discrepancies lie far below the
observational uncertainties and therefore they do not
affect  the comparison with observational data. In computing
the neutron star configurations, we have checked the
violations of the general-relativistic virial identities
GRV2 \citep{bonazzola1973, bonazzolagourgoulhon1994} and
GRV3 \citep{gourgoulhonbonazzola1994}.  The absolute
deviations lie between $10^{-3}$ and $10^{-6}$ thus
confirming the high precision of the analytical
representation of the EoSs.

%%%%%%%%%%%%%%%%%%%%%%%%%%%%%%%%%%%%%%%%%%%%%%%%%%%%%%%%%%%%%%%%%%%%%%%%
\section{Conclusions}
\label{sect:concl}

We constructed analytical representations, in terms of
the continuous and differentiable functions of a single
chosen variable, of three recently developed equations of
state BSk19, BSk20, and BSk21
\citep{PearsonGC11,Pearson-ea12}. We considered two
choices of the independent variable. The first one is
the mass density $\rho$.
Then \req{fit.P} gives function $P(\rho)$, which fits the
numerical EoS tables at
$10^6~\gcc<\rho\lesssim2\times10^{16}~\gcc$ with a typical
error of $\sim1$\%. The baryon number density $\nb$ can be
calculated from \req{Integral} to satisfy exactly the first
law of thermodynamics. Alternatively, $\nb$ can be evaluated
using our fit (\ref{fit.n-rho}). A variant is to choose
$\nb$ as an independent variable and calculate $\rho(\nb)$ 
from the fit (\ref{fit.rho-n}) and $P(\rho(\nb))$ from
\req{fit.P}. Then, if necessary, $\rho$ can be corrected
using the first integral relation in \req{Integral}.

\begin{figure}
\begin{center}
\includegraphics[width=\columnwidth]{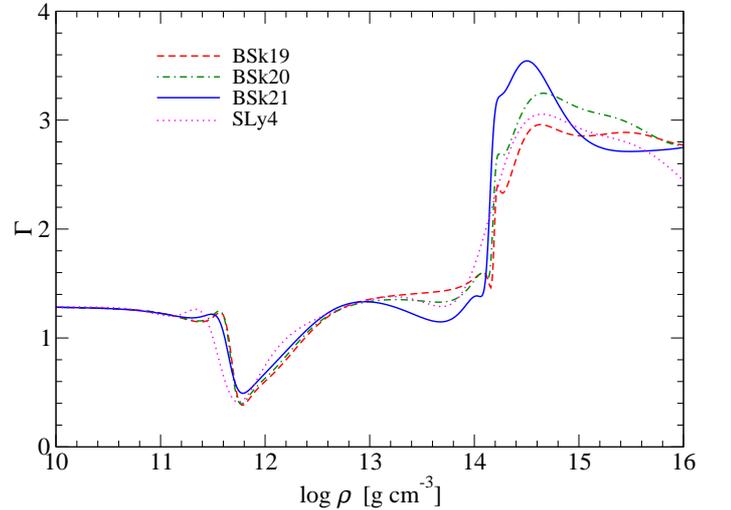}
\end{center}
\caption{Adiabatic index $\Gamma$ for SLy4 and BSk EoSs.
}
\label{fig:Gamma}
\end{figure}

Differentiation of
$P(\rho)$ then yields analytical representations of the
adiabatic index
\begin{equation}
   \Gamma=\frac{\nb}{P}\,\frac{\dd P}{\dd \nb}
   = \left[ 1+ \frac{P}{\rho c^2} \right]
   \frac{\rho}{P}\frac{\dd P}{\dd\rho}
~,
\end{equation}
which is included in the computer code that realizes the
fit. Different regions of neutron-star interior are
characterized by distinct behavior of $\Gamma$ as discussed,
e.g., in \citet{HP04}. This behavior remains qualitatively
the same for different EoSs, but quantitative differences
can be significant, as illustrated in Fig.~\ref{fig:Gamma}.

%%%%%%%%%%%%%%%%%%%%%%%%%%%%%%%%%%%%%%%%%%%%%%%%%%%%%%%%%%%%%%%%%%%%%%%%
%%%%%%%%%%%%%%%%%%%%%%%%%%%%%%%%%%%%%%%%%%%%%%%%%%%%%%%%%%%%%%%%%%%%%%%%

The other choice of the independent variable is the
pseudo-enthalpy $H$, \req{H.def}. This choice is
particularly advantageous for simulations of neutron-star
dynamics. We represented the EoSs by the continuous and
differentiable functions $P(H)$, $\rho(H)$, and $n(H)$,
where $\rho(H)$ is given by \req{fit.rho-h} with typical
accuracy $\sim1$\%, while $P(H)$ and $\nb(H)$ are calculated
from the functions $P(\rho)$ and $\nb(\rho)$, respectively. 

We also obtained analytical representations of number
fractions of neutrons, protons, electrons and muons in the
inner crust and the core, and nuclear shape parameters in
the inner crust of a neutron star as functions of $\nb$.
These results can be used, e.g., for neutron-star cooling
simulations. As an application, we calculate electron
conductivities in the crust with the use of the BSk models.
Compared to the results of the smooth-composition model
\citep{GYP}, we find that the BSk models yield appreciably
higher electrical conductivities in the inner crust of a
neutron star.

Finally, we estimated the errors introduced in the fitting 
formulae on global neutron-star properties and showed that
they lie far below the observational uncertainties and
therefore they do not affect the comparison of theoretical
models with observational data.

The present work is mainly
aimed at astrophysicists as they do not have to perform
nuclear physics calculations for simulations of neutron-star
structure and evolution.
When required, the same method of constructing analytical
approximations can be applied to other EoSs (see,
e.g., \citealp{HPY} and \citealp{lattimer2012}, for
references).

\begin{acknowledgements}
The work of A.Y.P.\ was partially supported by the Ministry
of Education and Science of the Russian Federation
(Agreement No.\,8409, 2012), the Russian Foundation for
Basic Research (RFBR grant 11-02-00253-a), the Russian
Leading Scientific Schools program (grant NSh-4035.2012.2),
and the Programme National de Physique Stellaire (CNRS/INSU,
France). The work of A.F.F., N.C.\ and S.G.\ was supported
by FNRS (Belgium).  The work of J.M.P.\ was supported by the
NSERC (Canada). 
\end{acknowledgements}

%%%%%%%%%%%%%%%%%%%%%%%%%%%%%%%%%%%%%%%%%%%%%%%%%%


\begin{thebibliography}{}

\bibitem[Aguilera et al.(2009)]{Aguilera-ea09}
Aguilera, D. N., Cirigliano, V., Pons, J. A., Reddy, S., \& Sharma, R.
2009,
\prl, 102, 091101

\bibitem[Akmal et al.(1998)]{APR}
Akmal, A., Pandharipande, V. R., \& Ravenhall D. G.
1998,
\prc, 58, 1804

\bibitem[Audi et al.(2003)]{audi2003}
Audi, G., Wapstra, A.~H., \& Thibault, C. 
2003, Nucl. Phys. A, 729, 337 

\bibitem[Baym et al.(1971)]{bps1971}
Baym G., Pethick, C., \& Sutherland P. 
1971, 
\apj, 170, 299

\bibitem[Bisnovatyi-Kogan \& Romanova(1982)]{BisnoRomanova}
Bisnovatyi-Kogan, G. S., \& Romanova, M. M.
1982,
Sov. Phys. JETP, 56, 243

\bibitem[Bonazzola(1973)]{bonazzola1973}
Bonazzola, S.
1973, 
\apj, 182, 335

\bibitem[Bonazzola et al.(1993)]{Bonazzola-ea93}
Bonazzola, S., Gourgoulhon, E., Salgado, M., \& Marck, J.~A.
1993,
\aap, 278, 421

\bibitem[Bonazzola \& Gourgoulhon(1994)]{bonazzolagourgoulhon1994}
Bonazzola, S., \& Gourgoulhon, E.
1994, 
Class.\ Quantum Grav., 11, 1775 

\bibitem[Carbone et al.(2010)]{carbone2010}
Carbone, A., Col{\`o}, G., Bracco, A., et al.
2010,
\prc, 81, 041301

\bibitem[Cackett et al.(2010)]{Cackett-ea10}
Cackett E. M., Brown E. F., Cumming, A., et al.
2010,
\apj, 722, L137

\bibitem[Centelles et al.(2009)]{centelles2009}
Centelles, M., Roca-Maza, X., Vi{\~n}as, X., \& Warda, M. 
2009, 
\prl, 102, 122502 

\bibitem[Chamel(2010)]{Chamel10}
Chamel, N.
2010,
\prc, 82, 014313

\bibitem[Chamel(2012)]{Chamel12}
Chamel, N.
2012,
\prc, 85, 035801

\bibitem[Chamel et al.(2009)Chamel, Goriely, \& Pearson]{ChamelGP09}
Chamel, N., Goriely, S., \& Pearson, J.~M.
2009,
\prc, 80, 065804

\bibitem[Chamel et al.(2011)]{Chamel-ea11}
Chamel, N., Fantina, A.~F., Pearson, J.~M., \& Goriely, S.
2011,
\prc, 84, 062802(R)

\bibitem[Chen et al.(2009)]{chen2009}
Chen, L.-W., Cai, B.-J., Ko, C.~M., et al. 
2009, 
\prc, 80, 014322 

\bibitem[Chen et al.(2010)]{chen2010}
Chen, L.-W., Ko, C. M., Li, B.-A., \& Xu, J.
2010, 
\prc, 82,  024321

\bibitem[Chugunov(2012)]{Chugunov12}
Chugunov, A. I.
2012,
Astron.~Lett., 38, 25

\bibitem[Chugunov \& Haensel(2007)]{ChugunovHaensel07}
Chugunov, A. I., \& Haensel, P.
2007,
\mnras, 381, 1143

\bibitem[Col\`o et al.(2004)]{colo04}
Col{\`o}, G., van Giai, N., Meyer, J., Bennaceur, K., \& Bonche, P.
2004, 
\prc, 70, 024307 

\bibitem[Danielewicz et al.(2002)]{danielewicz2002}
Danielewicz, P., Lacey, R., \& Lynch, W.~G. 
2002, 
Science, 298, 1592 

\bibitem[Danielewicz \& Lee(2009)]{danielewiczlee2009}
Danielewicz, P., \& Lee, J.
2009,
Int. J. Mod. Phys. E, 18, 892
 
\bibitem[Daoutidis \& Goriely(2011)]{daoutidisgoriely2011}
Daoutidis, I., \& Goriely, S.
2011,
\prc, 84, 027301 

\bibitem[Douchin \& Haensel(2000)]{DouchinHaensel00}
Douchin, F., \& Haensel, P., 
2000,
Phys. Lett. B, 485, 107

\bibitem[Douchin \& Haensel(2001)]{DouchinHaensel01}
Douchin, F., \& Haensel, P., 
2001,
\aap, 380, 151

\bibitem[Dutra et al.(2012)]{dutra2012}
Dutra, M., Louren{\c c}o, O., S{\'a} Martins, J.~S., et al.
2012, 
\prc, 85, 035201 

\bibitem[Fantina et al.(2012)]{Fantina-ea12}
Fantina, A. F., Chamel, N., Pearson, J. M., \& Goriely, S.
2012,
J. Phys.: Conf. Ser., 342, 012003

\bibitem[Friedman \& Pandharipande(1981)]{FP81}
Friedman, B., \& Pandharipande, V. R.
1981,
Nucl. Phys. A, 361, 502

\bibitem[Gale et al.(1990)]{Gale-ea90}
Gale, C., Welke, G. M., Prakash, M., Lee, S. J., \& Das Gupta, S.
1990,
\prc, 41, 1545

\bibitem[Gnedin et al.(2001)]{GYP}
Gnedin, O.~Y., Yakovlev, D.~G., \& Potekhin, A.~Y.
2001,
\mnras, 324, 725

\bibitem[Goriely et al.(2010)]{Goriely-ea10}
Goriely, S., Chamel, N., \& Pearson, J. M.
2010,
\prc, 82, 035804

\bibitem[Goriely et al.(2013)]{Goriely-ea13}
Goriely, S., Chamel, N., \& Pearson, J. M.
2013,
\prc, 88, 024308

\bibitem[Gourgoulhon \& Bonazzola(1994)]{gourgoulhonbonazzola1994}
Gourgoulhon, E., \& Bonazzola, S.
1994, 
Class. Quantum Grav., 11, 443 

\bibitem[Haensel(1995)]{Haensel95}
Haensel, P.
1995,
\ssr, 74, 427

\bibitem[Haensel \& Pichon(1994)]{HP94}
Haensel, P., \& Pichon, 
1994,
\aap, 283, 313

\bibitem[Haensel \& Potekhin(2004)]{HP04}
Haensel, P., \& Potekhin, A.~Y.
2004,
\aap, 498, 191

\bibitem[Haensel \& Proszynski(1982)]{HaenselProsz82}
Haensel, P., \&  Proszynski, M. 
1982, 
\apj, 258, 306

\bibitem[Haensel \& Zdunik(1990)]{HZ90}
Haensel, P., \& Zdunik, J.~L. 
1990,
\aap, 229, 117

\bibitem[Haensel \& Zdunik(2003)]{HZ03}
Haensel, P., \& Zdunik, J.~L. 2003,
\aap, 404, L33

\bibitem[Haensel et al.(2007)Haensel, Potekhin, \& Yakovlev]{HPY}
 Haensel, P., Potekhin, A.~Y., \& Yakovlev, D.~G.
 2007,
Neutron Stars 1: Equation of State and Structure
(New York: Springer)

\bibitem[Hameury et al.(1983)Hameury, Heyvaerts, \& Bonazzola]{HameuryHB83}
Hameury, J. M., Heyvaerts, J., \& Bonazzola, S.
1983,
\aap, 121, 259

\bibitem[Hessels et al.(2006)]{Hessels-ea06}
Hessels, J. W. T., Ransom, S. M., Stairs, I. H., et al.
% Freire P.C.C., Kaspi V.M., Camilo, F., 
2006,
Science, 311, 1901

\bibitem[Hughto et a.(2011)]{Hughto-ea11}
Hughto, J., Schneider, A. S., Horowitz, C. J., \& Berry, D. K.
2011,
\pre, 84, 016401

\bibitem[Kaminker et al.(1999)]{Kam-ea99}
Kaminker, A.~D., Pethick C.~J., Potekhin A.~Y., Thorsson V., \& Yakovlev D.~G.
1999,
\aap, 343, 1009

\bibitem[Kl\"ahn et al.(2006)]{Klaehn-ea06}
Kl\"ahn, T., Blaschke, D., Typel, S., et al.
2006,
\prc, 74, 035802

\bibitem[Lattimer(2012)]{lattimer2012}
Lattimer, J.~M. 
2012, 
Annu. Rev. Nucl. Particle Sci., 62, 485 

\bibitem[Lattimer \& Lim(2013)]{lattimerlim2013}
Lattimer, J.~M., \& Lim, Y.
2013,
\apj, 771, 51 

\bibitem[Lattimer et al.(1991)]{Lattimer-ea91}
Lattimer, J. M., Pethick, C. J., Prakash, M., \& Haensel, P.
1991,
\prl, 66, 2701

\bibitem[Li \& Schulze(2008)]{LiSchulze}
Li, Z. H., \& Schulze, H. J.
2008,
\prc, 78, 028801

\bibitem[Onsi et al.(2008)]{Onsi-ea08}
Onsi, M., Dutta, A. K., Chatri, H., et al.{}
% Goriely, S., Chamel, N., \& Pearson, J. M.
2008,
\prc, 77, 065805

\bibitem[Oyamatsu(1993)]{Oyamatsu93}
Oyamatsu, K.
1993,
Nucl.~Phys. A, 561, 431

\bibitem[Page et al.(2006)Page, Geppert, \& Weber]{PageGW}
Page, D., Geppert, U., \& Weber, F.
2006,
Nucl.~Phys. A, 777, 497

\bibitem[Pandharipande \& Ravenhall(1989)]{Pandharipande89}
Pandharipande, V.~R., \& Ravenhall, D.~G. 
1989,
in Nuclear Matter and Heavy Ion Collisions,
NATO ADS Ser., vol.~B205, 
ed. M.~Soyeur, H.~Flocard, B.~Tamain, \& M.~Porneuf
(Dordrecht: Reidel), 103

\bibitem[Pearson et al. (2009)]{Pearson2009}
Pearson, J. M., Goriely, S., Chamel, N., Samyn, M., \& Onsi, M.
2009, in
Bulk Nuclear Properties: 5th ANL/MSU/JINA/INT FRIB Workshop,
AIP Conf. Proc.,
ed. P.~Danielewicz,
1128, 29
% vol.~1128 
% (Melville, NY: AIP), 29

\bibitem[Pearson et al.(2011)Pearson, Goriely, \& Chamel]{PearsonGC11}
Pearson, J.~M., Goriely, S., \& Chamel, N.
2011,
\prc, 83, 065810

\bibitem[Pearson et al.(2012)]{Pearson-ea12}
Pearson, J.~M., Chamel, N., Goriely, S., \& Ducoin, C.
2012,
\prc, 85, 065803

\bibitem[P\'erez-Azorin et al.(2006)]{Perez-ea06}
P\'erez-Azorin, J. F., Miralles, J. A., \& Pons, J. A.
2006,
\aap, 451, 1009

\bibitem[Piekarewicz et al.(2012)]{piekarewicz2012}
Piekarewicz, J., Agrawal, B. K., Col{\`o}, G., et al. 
2012,
\prc, 85, 041302 

\bibitem[Pons et al.(2009)Pons, Miralles, \& Geppert]{PonsMG}
Pons, J. A.,  Miralles, J. A., \& Geppert, U.
2009,
\aap, 496, 207

\bibitem[Potekhin \& Yakovlev(2001)]{PotekhinYakovlev01}
Potekhin, A. Y., \& Yakovlev, D. G.
2001,
\aap, 374, 213

\bibitem[Potekhin et al.(1999)]{Potekhin-ea99}
Potekhin, A. Y., Baiko, D. A., Haensel, P., \& Yakovlev, D. G.
1999,
\aap, 346, 345

\bibitem[Reinhard \& Nazarewicz(2010)]{reinhardnazarewicz2010}
Reinhard, P. G. \& Nazarewicz, W.
2010,
\prc, 81, 051303 

\bibitem[Rogers et al.(1996)Rogers, Swenson, \& Iglesias]{OPAL-EoS}
Rogers, F.~J., Swenson, F.~J., \& Iglesias, C.~A.
1996,
\apj, 456, 902

\bibitem[Shternin \& Yakovlev(2006)]{ShterninYakovlev06}
Shternin, P. S., \& Yakovlev, D. G.
2006,
\prd, 74, 043004

\bibitem[Shternin et al.(2007)]{Shternin-ea07}
Shternin, P. S., Yakovlev, D. G., Haensel, P., \& Potekhin, A. Y.
2007,
\mnras, 382, L43

\bibitem[Steiner \& Gandolfi(2012)]{steinergandolfi2012}
Steiner, A.~W., \& Gandolfi, S. 
2012, 
\prl, 108, 081102 

\bibitem[Stergioulas(2003)]{Sterg03}
Stergioulas, N. 2003, Living Rev.\ Relativity, 6, 3,
URL http://www.\linebreak[1]livingreviews.\linebreak[1]org/lrr-2003-3/
 
\bibitem[Trippa et al.(2008)]{trippa2008}
Trippa, L., Col{\`o}, G. \& Vigezzi, E. 
2008,
\prc, 77, 061304

\bibitem[Tsang et al.(2012)]{tsang2012}
Tsang, M.~B., Stone, J.~R., Camera, F., et al. 
2012, 
\prc, 86, 015803 

\bibitem[Tsang et al.(2009)]{tsang2009}
Tsang M.~B., Zhang, Y., Danielewicz, P., et al. 2009, 
\prl, 102, 122701

\bibitem[Yakovlev \& Pethick(2004)]{YakPethick}
Yakovlev, D.~G., \& Pethick, C.~J.
2004,
\araa, 42, 169

\bibitem[Yakovlev et al.(2008)]{Yak-ea08}
Yakovlev, D. G., Gnedin, O. Y., Kaminker, A. D., \& Potekhin, A. Y.
2008, 
in
40 Years of Pulsars: Millisecond Pulsars, Magnetars and More, 
AIP Conf. Proc.,
eds. C.~Bassa, Z.~Wang, A.~Cumming, \& V.~M. Kaspi,
983, 379
% vol.~983,
% (Melville, NY: AIP), 379

\bibitem[Warda et al.(2009)]{warda2009}
Warda, M., Vi{\~n}as, X., Roca-Maza, X., \& Centelles, M. 
2009, 
\prc, 80, 024316 

\end{thebibliography}
\end{document}